\documentclass[preprint,11pt]{elsarticle}

\usepackage[english]{babel}
\usepackage{amssymb} 
\usepackage{amsmath}
\usepackage{bm}
\usepackage{url}
\usepackage[hidelinks]{hyperref}
 
\usepackage{booktabs} 
\usepackage{subcaption}
\usepackage{siunitx}
\usepackage{xcolor}
\usepackage{listings}
\usepackage{subfiles}
\usepackage{array}
\usepackage{enumitem}
\usepackage{longtable}
\usepackage{tablefootnote}

\DeclareSIUnit\angstrom{\text{\AA}}

\newcounter{bla}

\journal{Computer Physics Communications}

\begin{document}

\begin{frontmatter}

\title{BzScope: an absolute cross section calculator for neutron-phonon scattering}

\author[a,b]{Ming Tang}
\author[a,b]{Zi-Yi Pan}
\author[a,b,c]{Ni Yang}
\author[d]{Thomas Kittelmann}
\author[a,b]{Xiao-Xiao Cai\corref{author}}

\cortext[author] {Corresponding author.\\\textit{E-mail address:} caixx@ihep.ac.cn}

\address[a]{Institute of High Energy Physics, Chinese Academy of Sciences, China}
\address[b]{Spallation Neutron Source Science Center, China}
\address[c]{University of Chinese Academy of Sciences, China}
\address[d]{European Spallation Source ERIC, Lund, Sweden}

\begin{abstract}

\texttt{BzScope} is a Python package designed for efficiently calculating absolute cross sections of neutron-phonon inelastic scattering  for crystalline powders in large phase spaces, addressing the limitations of traditional histogramming techniques in reproducing sharp structures and ensuring convergence. The package employs an adapted integral method and supports calculations of single- and two-phonon scattering functions in ideal crystalline powders, with numerical robustness up to a momentum transfer of \SI{100}{\angstrom^{-1}}. Higher order scatterings up to several hundred orders are calculated by incoherent approximation in a  well-established thermal neutron scattering physics package, NCrystal. In addition, a NCrystal plugin is made available for NCrystal-enabled Monte Carlo packages, facilitating direct comparison between the new physics and experimental data.

Validation against NCrystal demonstrates good agreement in incoherent scattering for cubic systems Ni. 
In addition, it shows improved accuracy for low-symmetry materials  \( \text{NiP}_2 \) by avoiding the isotropic atomic displacement approximations in NCrystal. 
Benchmarks the experimental  differential cross section of LiH  and total cross section of Be confirm its reliability. 

\texttt{BzScope} integrates with NCrystal via a plugin and therefore can be directly used in any NCrystal-enabled Monte Carlo package. This tool enhances the efficiency and accuracy of neutron scattering simulations, advancing the study of condensed matter dynamics.

\end{abstract}

\begin{keyword}
scattering function; neutron-phonon scattering; inelastic neutron scattering; numerical calculation; scattering cross section.
\end{keyword}

\end{frontmatter}


{\bf PROGRAM SUMMARY}

\begin{small}
\noindent
{\em Program Title:} \texttt{BzScope}                                        \\
{\em CPC Library link to program files:} (to be added by Technical Editor) \\
{\em Developer's repository link:} \url{https://code.ihep.ac.cn/cinema-developers/bzscope} \\
{\em Code Ocean capsule:} (to be added by Technical Editor)\\
{\em Licensing provisions(please choose one):} GPLv3 \\
{\em Programming language:} Python                                  \\
{\em Supplementary material:}                                 \\
{\em Journal reference of previous version:}*                  \\
{\em Does the new version supersede the previous version?:}*   \\
{\em Reasons for the new version:*}\\
{\em Summary of revisions:}*\\
{\em Nature of problem(approx. 50-250 words):} High precision calculation of absolute cross sections of neutron-phonon inelastic scattering in adequately large phase spaces is critical for the study of condensed matter dynamics, especially the calculation of the contributions from low-order phonon processes. However, traditional calculation techniques suffer from accuracy, resolution and the subsequent iterative parameter tuning. In addition, it is not straight-forward to include the calculated cross sections in a realistic Monte Carlo simulation to reproduce the experiment. \\
{\em Solution method(approx. 50-250 words):} A new calculation tool based on an adaptive integration method and harmonic approximation is proposed. It can calculate the coherent scattering functions of single- and double-phonon on the absolute scale in ideal crystalline powders with numerical robustness up to a momentum transfer of \SI{100}{\angstrom^{-1}}. Higher-order scattering up to a few hundred orders can be approximated by incoherent approximation. Hence a complete inelastic scattering function can be obtained.\\
{\em Additional comments including restrictions and unusual features (approx. 50-250 words):}
Cross section data can be exported to the NCrystal formatted data file by a provided top-level tool. 
Therefore the detailed low-order coherent phonon scattering physics can be used in any NCrystal-enabled Monte Carlo  package to simulate the inelastic neutron scattering experiment. 
The proposed tool supports force constants format that is compatible with the Phonopy format, making it convenient to cooperate with mainstream density functional theory (DFT) tools through Phonopy.
\\
   \\
\end{small}

\

\section{Introduction}

In recent years, Monte Carlo methods have become increasingly prevalent in the interpretation of neutron scattering data~\cite{kim1999}. These methods provide a robust framework for simulating complex physical processes by leveraging statistical sampling methods. 
Advanced geometry simulation techniques, such as the union component~\cite{mcstas_union} and the particle transport method~\cite{PAN2024109004}, accommodate a wide range of scattering phenomena, including complex scattering processes and multiple scattering events.

To accurately simulate the interactions, the advanced Monte Carlo method firstly calculates the integral cross section over a large phase space of the scattering function, then samples the scattering function to generate individual neutron scattering events~\cite{CAI2019400}. Each event is treated as a discrete interaction, and the cumulative response of these events can be used to model the detector's response. By generating a large number of simulated events, these methods can produce statistically significant results that closely mirror experimental observations.

Neutron inelastic scattering experiments are vital for probing the microscopic dynamics of matter, enabling direct measurements of phonons, spin excitations, and energy transfer processes critical to understanding condensed matter phenomena. 
The observed experimental data requires complementary calculations for proper interpretation.
However, it is generally challenging to calculate the overall scattering function of inelastic scattering with high resolution in a straightforward manner.

For example, one of the primary methods for computing single phonon scattering functions is the histogramming technique. This method involves generating a fixed grid in reciprocal space to represent phonon states and accumulating the contributions of each phonon in different Brillouin zones. 
The histogramming technique is essentially a form of numerical integration, specifically the trapezoid rule, which approximates the integral of a function by dividing the area under the curve into trapezoids and summing their areas.

While straightforward to implement, the histogramming technique has a number of limitations. In particular, it is challenging to reproduce sharp structures in the scattering phase space using this low-order method. Additionally, developing convergence tests is very challenging. Besides the rapid growth of the phonon grid size with increased sampling dimensions, it is also difficult to define convergence criteria across the entire scattering phase space as required by the sampling principle of Monte Carlo  simulation.

The present package \texttt{BzScope}, standing for \texttt{Brillouin Zone Scope}, utilises the adapted integral method to significantly increase the efficiency to calculate the inelastic scattering functions in a large phase space. 
In the current version, the integral for single and two phonon scatterings in ideal crystalline powders is implemented.
The Fourier interpolation method in the Euphonic package~\cite{Fair2022} is used to obtain phonon frequencies and polarization vectors. 
The Vegas algorithm~\cite{PETERLEPAGE1978192, LEPAGE2021110386} is employed to perform the powder averaging. To facilitate direct comparison of the calculated functions with experimental results in Monte Carlo simulations, the upper limit of the scattering function is typically close to 100\si{\angstrom^{-1}}, within which BzScope is numerically robust in all tested cases.
The scattering functions can be loaded into the thermal scattering engine NCrystal via a dedicated plugin~\cite{CAI2020106851}.

The scope of this work is to validate the numerical accuracy of the implementation and demonstrate the usefulness of the package stand-alone and as a plugin to NCrystal~\cite{CAI2020106851}. The calculated results are compared with experimental data. The validity of incoherent approximation is also computationally investigated.

\section{Theory}

\subsection{Scattering theory in brief}

In a system of interacting particles, the scattering process can be characterized by the correlations of particle pairs in space and in time~\cite{Hove1954,squiresnew,CAI2020106851,KITTELMANN2021108082}. 
In this section, we confine our introduction to neutron nuclear scattering in crystals.

In the case of measuring a large ensemble that contains $N$ statistically equivalent scatterers, where each equivalent nucleus can be indexed by $j$, the intermediate function $I$ is

\begin{equation}
\label{eIg}
I(\bm{Q}, t) =  \frac{1}{2\pi\hbar}  \sum_{j^\prime, j }  
\overline{b_{j^\prime} {b}_{j}}
\langle
\exp[-i \bm{Q} \cdot \bm{R}_{j^\prime}(0)]
\exp[i \bm{Q} \cdot \bm{R}_j(t)]
\rangle 
\end{equation}

Here $b_j$ is the scattering length of the $j$-th nucleus. The overline above the lengths denotes  averaging over all the scatterers.
$\bm{R}(t)$ is the operator of the nucleus position,
$\bm{Q}$ is the momentum transfer in units of $\hbar$.
If there is no correlation between the scattering length of different nuclei, the effective scattering length product of a distinct pair of nuclei, i.e. $j^\prime \neq j$, $ \overline{b_{j^\prime} {b}_{j}}$ then equals $ \overline{b}_{j^\prime}\overline{b}_j$. 
Similarly, a pair made of a single nuclei, i.e. $j^\prime = j$, $\overline{b_{j^\prime} {b}_{j}}$ is $ \overline{{b}_j^2}$. Hence  one can break down $I(\bm{Q}, t)$ into the main diagonal elements and the off-diagonal elements that respectively corresponds to self scattering and distinct functions.

\begin{multline}
\label{eI2}
I(\bm{Q}, t) =  \frac{1}{2\pi\hbar}  \sum_{\substack{j', j \\ j' \neq j}   }  
\overline{b}_{j^\prime}\overline{b}_j
\langle
\exp[-i \bm{Q} \cdot \bm{R}_{j^\prime}(0)]
\exp[i \bm{Q} \cdot \bm{R}_j(t)]
\rangle \\
+\frac{1}{2\pi\hbar}  \sum_{j}  
\overline{{b}_j^2}
\langle
\exp[-i \bm{Q} \cdot \bm{R}_{j}(0)]
\exp[i \bm{Q} \cdot \bm{R}_j(t)]
\rangle 
\end{multline}

Eq.~\ref{eI2} can be rearranged to represent contributions from the coherent and incoherent scattering functions as  
\begin{multline}
\label{eI3}
I(\bm{Q}, t) =  \frac{1}{2\pi\hbar}  \sum_{\substack{j', j}   }  
\overline{b}_{j^\prime}\overline{b}_j
\langle
\exp[-i \bm{Q} \cdot \bm{R}_{j^\prime}(0)]
\exp[i \bm{Q} \cdot \bm{R}_j(t)]
\rangle \\
+\frac{1}{2\pi\hbar}  \sum_{j}  
\left[\overline{{b}_j^2}-\overline{b}_{j}\overline{b}_j
\right]
\langle
\exp[-i \bm{Q} \cdot \bm{R}_{j}(0)]
\exp[i \bm{Q} \cdot \bm{R}_j(t)]
\rangle 
\end{multline}

Let the instantaneous atomic position be
\begin{equation}
\label{eE}
\bm{R}_{ld}(t) = \bm{l} + \bm{d} + \bm{u}_{ld}(t)
\end{equation}
where $\bm{l}$ and $\bm{d}$ are the lattice vector of a unit cell and the equilibrium position of a nucleus, respectively,  $\bm{u}$ is the instantaneous displacement from its equilibrium position. In a crystal, the coherent scattering, i.e. the first term of Eq.~\ref{eI3}, becomes

\begin{multline}
\label{ecoh1}
I_{\text{coh}}(\bm{Q}, t) 
= \frac{1}{2\pi\hbar}  \sum_{l^\prime d^\prime, ld } 
\overline{b}_{l^\prime d^\prime}\overline{b}_{ld}
\exp[i\bm{Q}\cdot (\bm{l} + \bm{d} -\bm{l^\prime}-\bm{d^\prime} ) ]
\\
\langle
\exp[-i \bm{Q} \cdot \bm{u}_{l^\prime d^\prime}(0)]
\exp[i \bm{Q} \cdot \bm{u}_{ld}(t)]
\rangle 
\end{multline}

If the crystal consists a large number of unit cells, each inner sum is almost equal. 
In that case, Eq.~\ref{ecoh1} can be simplified by setting the $\bm{l^\prime}=\bm{l}_o=[0, 0, 0]$ in the outer summation as the spatial origin 

\begin{multline}
\label{eI_beforeHA}
I_{\text{coh}}(\bm{Q}, t) 
=\frac{N}{2\pi\hbar} 
\sum_{l} 
\exp(i\bm{Q}\cdot \bm{l} )\\
\sum_{ d^\prime, d } 
\bar{b}_{d^\prime} \bar{b}_{d }
\exp[i\bm{Q}\cdot ( \bm{d} - \bm{d^\prime} ) ]
\langle
\exp[-i \bm{Q} \cdot \bm{u}_{\bm{l}_o d^\prime}(0)]
\exp[i \bm{Q} \cdot \bm{u}_{ld}(t)]
\rangle 
\end{multline}
here $N$ is the number of cells in the crystal.

For incoherent scattering, the scattering function originated from self-correlation, meaning $\bm{l}'=\bm{l}$ and $\bm{d}'=\bm{d}$. The intermediate scattering function is 
\begin{multline}
\label{eInco}
I_{\text{incoh}}(\bm{Q}, t) 
= \frac{1}{2\pi\hbar}  \sum_{d } 
\left[\overline{{b}_d^2}-\overline{{b}_{d}}^2
\right]
\langle
\exp[-i \bm{Q} \cdot \bm{u}_{d}(0)]
\exp[i \bm{Q} \cdot \bm{u}_{d}(t)]
\rangle 
\end{multline}

\subsection{Phonon expansion in large harmonic crystals}

In harmonic crystals, it has been shown that the thermal averaging obeys~\cite{Maradudin1962}.

\begin{equation}
\label{eThermalAvg}
\langle e^x e^y\rangle = \exp\big[\frac{1}{2} \langle x^2 + 2xy + y^2 \rangle\big]
\end{equation}
Therefore, the thermal averaging operator, in Eq.~\ref{eI_beforeHA}, can be expanded as 

\begin{multline}
\label{eHA}
\langle
\exp[-i \bm{Q} \cdot \bm{u}_{l^\prime d^\prime}(0)]
\exp[i \bm{Q} \cdot \bm{u}_{ld}(t)]\rangle 
= \exp\left(-\frac{1}{2} \langle (\bm{Q}\cdot \bm{u}_{ld})^2 \rangle \right)
\,
\exp\left(-\frac{1}{2} \langle (\bm{Q}\cdot \bm{u}_{l^\prime d^\prime})^2 \rangle \right)  \\
\exp\langle 
\bm{Q} \cdot \bm{u}_{l^\prime d^\prime}(0)  \bm{Q} \cdot \bm{u}_{ld}(t)
\rangle 
\end{multline}

Applying Eq.~\ref{eHA} to Eq.~\ref{eI_beforeHA} in the appendix, the coherent intermediate function becomes

\begin{align}
\label{eI4}
\begin{split}
I_{\text{coh}}(\bm{Q}, t) 
=&\frac{N}{2\pi\hbar} 
\sum_{l} 
\exp(i\bm{Q}\cdot \bm{l} )
\sum_{ d^\prime, d } 
\bar{b}_{l^\prime d^\prime} \bar{b}_{l d }
\exp[i\bm{Q}\cdot ( \bm{d} - \bm{d^\prime} ) ]
\\&
\exp(-\frac{1}{2} \langle [\bm{Q}\cdot \bm{u}_{d}]^2 \rangle)  
\exp(-\frac{1}{2} \langle [\bm{Q}\cdot \bm{u}_{d^\prime}]^2 \rangle)  
\exp\langle 
-\bm{Q} \cdot \bm{u}_{l_o d^\prime}(0)  \bm{Q} \cdot \bm{u}_{ld}(t)\rangle
\end{split}
\end{align}

Notice that in Eq.~\ref{eI4}, the factor $\exp\left(-\frac{1}{2} \langle [\bm{Q} \cdot \bm{u}_{d}]^2 \rangle \right)$ is no longer dependent on the unit cell. This is because, in a crystal, any nucleus can be equivalently found in the unit cell at an arbitrary cell $\bm{l}$. The exponent of this factor is referred as the  Debye-Waller exponent, $W_d$~\cite{Maradudin1962}.  

\begin{equation}
\label{eDebye-WallerExp}
W_d =  \frac{1}{2} \langle (\bm{Q}\cdot \bm{u}_{d})^2 \rangle = 
\bm{Q}^T\cdot \left[ \frac{\hbar}{4M_dN}\sum_{qs} \frac{ \bm{e}_{ds_i}({\bm{q}_i}) \otimes \bm{e}^*_{ds_i}({\bm{q}_i})}{\omega_s(\bm{q})} \langle 2n_s +1 \rangle \right] \cdot \bm{Q}
\end{equation}

Also, the last exponential factor in Eq.~\ref{eI4} can be expanded in a Taylor series. Such an expansion provides significant physical insights into the scattering processes. Specifically, the \(n\)-th term in the expansion represents the scattering involving \(n\) phonons.

\begin{align}
\label{ePhononExp}
\begin{split}
 \exp \langle[-\bm{Q} \cdot \bm{u}_{l_o d^\prime}(0)]
[ \bm{Q} \cdot \bm{u}_{ld}(t)]\rangle  = \sum_{n=0}^{\infty} \frac{1}{n!} 
\Big(\langle [-\bm{Q} \cdot \bm{u}_{l_o d^\prime}(0)
][\bm{Q} \cdot \bm{u}_{ld}(t)]\rangle\Big)^n
\end{split}
\end{align}

Following the displacement operator defined by the  Eq.4.26 in~\cite{Lovesey1984}, the correlation function $\langle \bm{Q} \cdot \bm{u}_{l_0 d'}(0) \, \bm{Q} \cdot \bm{u}_{ld}(t) \rangle$ can be expressed as  

\begin{multline}
\label{eSingleCorrelation}
\langle \bm{Q} \cdot \bm{u}_{l_0 d'}(0) \bm{Q} \cdot \bm{u}_{ld}(t) \rangle = \frac{1}{(NM_{d'})^{1/2}} \frac{1}{(NM_d)^{1/2}} \sum_{q's'} \sum_{qs} \left( \frac{\hbar}{2\omega_{s'}(\bm{q}')} \right)^{1/2} \\
\left( \frac{\hbar}{2\omega_{s}(\bm{q})} \right)^{1/2} \times (\bm{Q} \cdot \bm{e}_{d's'}({\bm{q}'})) (\bm{Q} \cdot \bm{e}_{ds}({\bm{q}})) \exp(i\bm{q}'\cdot\bm{l}_0) \exp(i\bm{q}\cdot\bm{l}) \\
\times \langle \left[ {\hat{a}}_{s'}(\bm q') + {\hat{a}}_{s'}^+ (-\bm q') \right] \left[ {\hat{a}}_{s}(\bm q) \exp( - i t \omega_{s}(\bm q) ) + {\hat{a}}_{s}^+ (-\bm q) \exp(  i t \omega_{s}(\bm q) )\right] \rangle
\end{multline}

In this context, \(\hat{a}_{s'}\) and \(\hat{a}_{s}^+\) denote phonon annihilation and creation operators, respectively. Non-zero expectation values occur only for \(\langle \hat{a}_{s'}(\bm{q'}) \hat{a}_{s}^+(\bm{q}) \rangle\) and \(\langle \hat{a}_{s'}^+(\bm{q'}) \hat{a}_{s}(\bm{q}) \rangle\). Specifically, \(\langle \hat{a}_{s'}(\bm{q'}) \hat{a}_{s}^+(\bm{q}) \rangle = (\langle n[\omega_{s}(\bm{q})] \rangle + 1) \delta_{s's} \delta_{\bm{q}' \bm{q}}\) and \(\langle \hat{a}_{s'}^+(\bm{q'}) \hat{a}_{s}(\bm{q}) \rangle = \langle n[\omega_{s}(\bm{q})] \rangle \delta_{s's} \delta_{\bm{q}' \bm{q}}\), where \(\langle n[\omega_{s}(\bm{q})] \rangle\) represents the average number of phonons in the mode with frequency \(\omega_{s}(\bm{q})\), and the Kronecker deltas \(\delta_{s's}\) and \(\delta_{\bm{q}' \bm{q}}\) ensure that the expectation values are non-zero only when the modes and wave vectors match.
Thus the correlation function is

\begin{align}
\label{eUV}
&\langle \bm{Q} \cdot \bm{u}_{l_0 d'}(0) \bm{Q} \cdot \bm{u}_{ld}(t) \rangle = \frac{\hbar}{N(M_{d'} M_d)^{1/2}} \sum_{qs} \frac{[\bm{Q} \cdot \bm{e}_{d's}({\bm{q}})] [\bm{Q} \cdot \bm{e}_{ds}({\bm{q}})]}{2\omega_{s}(\bm{q})} \\
&\times \exp(-i\bm{q}\cdot \bm{l}) 
\left[ (\langle n(\omega_{s}(\bm{q})) \rangle + 1) \exp(  i t \omega_{s}(\bm{q}) ) + \langle n(\omega_{s}(\bm{q})) \rangle \exp(-  i t \omega_{s}(\bm{q}) ) 
\right]
\end{align}

Similarly, the second order correlation function can be expressed as 
$$
\begin{aligned}
\label{eUv2}
&\langle \bm{Q} \cdot \bm{u}_{l_0 d'}(0) \bm{Q} \cdot \bm{u}_{ld}(t) \rangle^2 \\
&= \frac{\hbar^2}{N^2(M_{d'} M_d)}  \sum_{q_1s_1} \sum_{q_2s_2} \frac{[\bm{Q} \cdot \bm{e}_{d's_1}({\bm{q}_1})] [\bm{Q} \cdot \bm{e}_{ds_1}({\bm{q}_1})]}{2\omega_{s_1}(\bm{q}_1)} 
\exp(-i\bm{q}_1\cdot \bm{l}) \\
&\times \left[ (\langle n(\omega_{s_1}(\bm{q}_1)) \rangle + 1) \exp(  -i t \omega_{s_1}(\bm{q}_1) ) + \langle n(\omega_{s_1}(\bm{q}_1)) \rangle \exp(   i t \omega_{s_1}(\bm{q}_1) ) \right] \\
&\times \frac{[\bm{Q} \cdot \bm{e}_{d's_2}({\bm{q}_2})] [\bm{Q} \cdot \bm{e}_{ds_2}({\bm{q}_2})]}{2\omega_{s_2}(\bm{q}_2)} 
\exp(-i\bm{q}_2\cdot \bm{l}) \\
&\times \left[ (\langle n(\omega_{s_2}(\bm{q}_2)) \rangle + 1) \exp(  -i t \omega_{s_2}(\bm{q}_2) ) + \langle n(\omega_{s_2}(\bm{q}_2)) \rangle \exp(   i t \omega_{s_2}(\bm{q}_2) ) \right]
\end{aligned}
$$

\subsection{Coherent elastic and inelastic scattering}

Given the correlation functions, it is straight forward to obtain the scattering functions for each order.
The zeroth order, i.e. coherent elastic scattering, can be expressed as

\begin{align}
\label{eIcoh0}
\begin{split}
I_{\text{coh}, 0}(\bm{Q}, t) 
=&\frac{N}{2\pi\hbar} 
\sum_{l} 
\exp(i\bm{Q}\cdot \bm{l} )
\sum_{ d^\prime, d } 
\bar{b}_{l^\prime d^\prime} \bar{b}_{l d }
\exp[i\bm{Q}\cdot ( \bm{d} - \bm{d^\prime} ) ]
\\&
\exp(-W_d)  
\exp(-W_{d'})  \\
=& \frac{N}{2\pi\hbar} 
\sum_{l} 
\exp(i\bm{Q}\cdot \bm{l} )
\left| \sum_{ d } 
 \bar{b}_{d }
\exp[i\bm{Q}\cdot \bm{d} ]
\exp(-\frac{1}{2} \langle [\bm{Q}\cdot \bm{u}_{d}]^2 \rangle)  \right|^2
\end{split}
\end{align}

Given that $\int \exp(-i\omega t) dt =2\pi\delta(\omega)$ and $\sum_{l} 
\exp(i\bm{Q}\cdot \bm{l} ) = (2\pi)^3 \sum_{\bm{\tau}}  \delta(\bm{Q}-\bm{\tau})/v$, the corresponding scattering function is

\begin{align}
\label{eScoh0}
\begin{split}
S_{\text{coh}, 0}(\bm{Q}, \omega) =& \int I_{\text{coh}, 0}(\bm{Q}, t) \exp(-i\omega t) dt \\
=& \frac{N}{\hbar} 
\frac{(2\pi)^3}{v} 
\delta(\omega)
\sum_{\bm{\tau}}  \delta(\bm{Q}-\bm{\tau})
\left| \sum_{ d } 
 \bar{b}_{d }
\exp[i\bm{Q}\cdot \bm{d} ]
\exp(-W_d)  \right|^2 
\end{split}
\end{align}

Based on Eq.~\ref{eUV}, the intermediate scattering function for the process of phonon creation is 

\begin{align}
\label{eIC1}
\begin{split}
I_{\text{coh},+}(\bm{Q}, t) 
=&\frac{N}{2\pi\hbar}
\sum_{ d^\prime, d } 
\bar{b}_{l^\prime d^\prime} \bar{b}_{l d }
\exp[i\bm{Q}\cdot ( \bm{d} - \bm{d^\prime} ) ]
\exp(-W_d)  
\exp(-W_{d'}) \\
&\frac{\hbar}{N(M_{d'} M_d)^{1/2}}
\times 
\sum_{qs} 
\sum_{l} 
\exp[i(\bm{Q} - \bm{q})\cdot \bm{l} ]
\frac{[\bm{Q} \cdot \bm{e}_{d's}({\bm{q}})] [\bm{Q} \cdot \bm{e}_{ds}({\bm{q}})]}{2\omega_{s}(\bm{q})} \\
&(\langle n(\omega_{s}(\bm{q})) \rangle + 1) \exp(  i t \omega_{s}(\bm{q}) ) 
\end{split}
\end{align}

Using $\sum_{l} \exp[i(\bm{Q}-\bm{q}) \cdot\bm{l}] =\frac{(2\pi)^3}{v} \delta(\bm{Q}-\bm{q}-\bm{\tau})
$, the expression can be simplified.
\begin{align}
\label{eIC2}
\begin{split}
I_{\text{coh},+}(\bm{Q}, t) 
=& \frac{1}{2\pi} \frac{(2\pi)^3}{v} 
\sum_{qs} 
\sum_{\bm{\tau}}  \sum_{ d } 
\frac{1}{2\omega_{s}(\bm{q})}\\
&\left|
\bar{b}_{d} 
\exp(i\bm{Q}\cdot  \bm{d}) 
\exp(-W_d)  \frac{\bm{Q} \cdot \bm{e}_{ds}({\bm{q}})}{M_{d}^{1/2}}
\right| ^2\\
&\delta(\bm{Q}-\bm{q}-\bm{\tau})
(\langle n(\omega_{s}(\bm{q})) \rangle + 1) \exp(  i t \omega_{s}(\bm{q}) )
\end{split}
\end{align}
 
Because
$\int \exp(-i[\omega_s(\bm{q})-\omega] t) dt =2\pi\delta[\omega-\omega_s(\bm{q})]$, the Fourier transform of the intermediate scattering function, the scattering function of single phonon creation, is
\begin{align}
\label{eS+}
\begin{split}
S_{\text{coh},+}(\bm{Q}, \omega)=& \int I_{\text{coh},+}(\bm{Q}, t) \exp (-i\omega t) d t\\
=& \frac{(2\pi)^3}{v} 
\sum_{qs} 
\sum_{\bm{\tau}}  \sum_{ d } 
\frac{1}{2\omega_{s}(\bm{q})}\\
&\left|
\bar{b}_{d} 
\exp(i\bm{Q}\cdot  \bm{d}) 
\exp(-W_d)   \frac{\bm{Q} \cdot \bm{e}_{ds}({\bm{q}})}{M_{d}^{1/2}}
\right| ^2\\
&(\langle n(\omega_{s}(\bm{q})) \rangle + 1) 
\delta( \omega - \omega_{s} )
\delta(\bm{Q}-\bm{q}-\bm{\tau})
\end{split}
\end{align}

Similarly, from Eq.~\ref{eInco} and Eq.~\ref{eSingleCorrelation}, 
the  incoherent scattering function for single phonon creation is

\begin{equation}
\label{eIncohSingleUp}
   S_{\text{incoh},+}(\bm{Q}, \omega)=\sum_d \frac{\overline{{b}_d^2}-\overline{{b}_{d}}^2}{2M_d} \exp(-2W_d) \sum_s \frac{| \bm{Q} \cdot \bm{e}_{ds}({\bm{q}})| ^2}{\omega_s} (\langle n(\omega_{s}(\bm{q})) \rangle + 1) \delta(\omega - \omega_s)
\end{equation}

For the mixture of $n_+$ creation and $n_-$ annihilation scatterings that involves $n$ phonons in total, it is straight forward to write the scattering functions as

\begin{align}
\label{eSntot}
\begin{split}
S_{\text{coh},{n}}
(\bm{Q}, \omega)
&=  \sum_{n_-=0}^n \frac{(2\pi)^3}{v}
\sum_{q_1 s_1, q_2 s_2, \ldots, q_n s_n}
\sum_{\bm{\tau}}  \sum_{ d } 
\prod_{i=1}^{n} \frac{1}{2\omega_{s_i}(\bm{q}_i)}
\frac{1}{n!}
\frac{\hbar^{(n-1)}}{N^{(n-1)}} \\
&\times \left|
\bar{b}_{d} 
\exp(i\bm{Q}\cdot  \bm{d}) 
\exp(-W_d) 
\frac{
\prod_{i=1}^{n_+} \bm{Q} \cdot \bm{e}_{ds_i}({\bm{q}_i})
\prod_{i'=1}^{n_-} \bm{Q} \cdot \bm{e}_{ds_{i'}}^*({\bm{q}_{i'}})
}
{M_{d}^{n/2}}
\right| ^2\\
&\times \prod_{i=1}^{n_+} (\langle n(\omega_{s_i}(\bm{q}_i)) \rangle + 1) 
\prod_{i'=1}^{n_-} \langle n(\omega_{s_{i'}}(\bm{q}_{i'})) \rangle\\ 
&\times \delta\left( \omega - \sum_{i=1}^{n_+} \omega_{s_i} + \sum_{i'=1}^{n_-} \omega_{s_{i'}} \right)
\delta\left(\bm{Q} - \sum_{i=1}^{n_+} \bm{q}_i 
+  \sum_{i'=1}^{n_-} \bm{q}_{i'}
- \bm{\tau} \right)
\end{split}
\end{align}

When $n$ is zero, the sum over an empty set is zero, and the product over an empty set is one. Hence, this equation is also valid for elastic scattering.

\subsection{The cubic Bravais incoherent model}
\label{ssCubicIncoAppr}

For cubic Bravais lattice, the atomic displacement is isotropic. The Debye-Waller exponent in Eq.~\ref{eDebye-WallerExp} can be expressed as~\cite{CAI2019400}
\begin{equation}
\label{eDebye-wallVDOS}
W=\frac{\hbar^{2} Q^{2}}{4 M} \int \frac{1}{\omega} \coth\left(\hbar \omega / 2 k_{b} T\right) \rho(\hbar \omega) \mathrm{d} \hbar \omega
\end{equation}

Where  $\rho$, the probability density of phonon energy, is known as the vibrational density of states (VDOS).

\begin{equation}
\label{eDOSomega}
\rho(\hbar \omega)=\frac{1}{N} \sum_{s} \delta\left[\hbar \omega-\hbar \omega_s\right]
\end{equation}

It has been shown that the incoherent scattering function in this case is~\cite{CAI2019400}

\begin{equation}
\label{eIncoherentSum}
S_{\text{incoh},{n}}
(Q, \omega)
= e^{-2W} \frac{(2W)^n}{n!} H_n(\hbar \omega)
\end{equation}

Where

\begin{equation}
\label{eHnFunc}
H_{n}(\hbar \omega)=\left\{\begin{array}{ll}
\delta(\hbar \omega) & , \quad n=0 \\
{\left[2 \hbar \omega \gamma(0) \sinh \left(\hbar \omega / 2 k_{b} T\right) \exp \left(\hbar \omega / 2 k_{b} T\right)\right]^{-1} \rho(\hbar \omega)} & , \quad n=1 \\
\int H_{1}\left(\hbar \omega^{\prime}\right) H_{N-1}\left(\hbar \omega-\hbar \omega^{\prime}\right) \mathrm{d} \hbar \omega^{\prime} & , \quad n>1
\end{array}\right.
\end{equation}

with 
\begin{equation}
\label{eGamma}
\gamma(t)=\int_{0}^{\infty} \frac{\rho(\hbar \omega)}{\omega}\left[\coth\left(\frac{\hbar \omega}{2 k_{b} T}\right) \cos (\omega t)+\mathrm{i} \sin (\omega t)\right] \mathrm{d} \omega
\end{equation}

The incoherent scattering function becomes polarization independent in this isotropic case.  
Owing to its simplicity, this model gains popularity in many numerical implementations, including NCrystal, even for non-cubic and non-Bravais lattice. In that case, the projected density of states (PDOS) are re-normalised and used in the place of the VDOS, as if the total incoherent scattering function were a linear combination of a series of cubic Bravais lattices that each are formed by one atom.

\begin{equation}
\label{ePDOSomega}
\rho_d(\hbar \omega)=\frac{1}{N} \sum_{s} \delta\left[\hbar \omega-\hbar \omega_s\right] 
\left | \bm{e}_{ds}({\bm{q}}) \right |^2 
\end{equation}

BzScope can explicitly calculate the incoherent scattering function based on no more than the harmonic approximation.
It is of interest to us to verify the validity of such an incoherent approximation in NCrystal. Further discussion can be found in section \ref{ssINcoh}.

\section{Implementation and top-level tools}

\subsection{Implementation}

\texttt{BzScope} relies on two well-established open source packages: Euphonic~\cite{Fair2022} and VEGAS~\cite{PETERLEPAGE1978192,LEPAGE2021110386}, to respectively solve for phonon frequencies and polarization directions from dynamic matrix, and implement for the high-dimensional adapted numerical integration.

Facilitated by the Euphonic package, the force constants in the popular phonopy~\cite{Togo_2023} format can be used to composite dynamic matrix. 
Hence, enabling the interface between \texttt{BzScope} and most of the mainstream density functional theory calculation packages, including VASP~\cite{Hafner1997} and Quantum Espresso~\cite{Giannozzi_2009} that are used in the validation section of this manuscript. 

The numerical implementation of \texttt{BzScope} leverages the properties of the dynamic matrix arrangement provided by the underlying Euphonic package that \(\omega_{s}(\bm{q}) = \omega_{s}(\bm{Q})\) and \(\bm{e}_{ds_i}({\bm{q}_i}) = \bm{e}_{ds_i}({\bm{Q}_i})\).
To further reduce computational demands, only neutron down-scattering, i.e., neutron energy loss, is calculated. The up-scattering component of the function is recovered by applying the detailed balance condition to the down-scattering results afterwards. 

The VEGAS+ adapted integration method is capable of  dynamically adjusting the sampling density to focus on regions of the integrand that contribute the most to the integral. 
Thanks to this method, the final numerical results include both the estimated mean and variances.

Considering the properties of the least symmetric crystals, the azimuthal angle \(\phi\) is restricted to the range \([0, \pi]\) to confine the problem to a hemisphere.
In addition, to ensure compatibility with the cross-section definition in NCrystal~\cite{CAI2020106851}, the final calculated scattering functions are converted to units of barns per atom.

There are mainly three types of classes in \texttt{BzScope} package, namely,  evaluators, integrands and workers.

\subsubsection{Evaluators}

Evaluators evaluate certain equations at sample points in three or six dimensions.

For the three-dimensional problems, a random point sampled in the 3D cube \([-0.5, 0.5]^3 \subset \mathbb{R}^3\) is sampled uniformly.
There are two concrete  classes in this type of problem.
The \texttt{PhononForDW} class evaluates the phonon contributions to each matrix element of the Debye-Waller exponent using Eq.~\ref{eDebye-WallerExp}. 
The \texttt{CoherentSinglePhonon} class evaluates phonon contributions to the coherent single phonon scattering function in Eq.~\ref{eSntot}. 

For the six-dimensional problems, two random reciprocal points are sampled representing the $\bm Q$ for the scattering and $\bm q$ for one phonon.  
There are two concrete inherited classes in this type of problem, as well.
The \texttt{CoherentTwoPhonon} class evaluates the coherent two phonons contribution using Eq.~\ref{eSntot} to the scattering function. 
To satisfy the momentum conservation, the correlated $\bm q$ point is translated into the correct Brillouin zone for the calculation.   
\texttt{IncoherentSinglePhonon} evaluates the incoherent single phonon contribution in Eq.~\ref{eInco} to the scattering function. In this case, without the delta function for moment conservation, $\bm Q$ and $\bm q$ are sampled without correlation.  
Notice that, this evaluator is developed for benchmarking the precision of this package against the proven incoherent scattering model in NCrystal. It is expected that they are numerically equivalent for cubic Bravais crystals.

\subsubsection{Integrands}

The ``integrand''s of the system defines the domain of the integration for a given evaluator.
Two integrands, 
\texttt{PhononPowderAverageIntegrand} and
\texttt{DebyeWallerIntegrand}, are inherited  from the BatchIntegrand class of the \texttt{vegas} package. They define the integration space for the \texttt{vegas} integrator based on the class type of related evaluator.
This type of classes also passes the $\bm Q$ and  $\bm q$ points sampled by the kernel of \texttt{vegas} to an evaluator.

For the \texttt{PhononPowderAverageIntegrand} integrand, there are two calculation methods. The $\bm q$ points are sampled in Cartesian coordinates, while the $\bm Q$ points are sampled in  spherical coordinates on either a concentric spherical shell in Eq.~\ref{eQpowder1} or a  spherical surface geometry in Eq.~\ref{eQpowder2}. 
The purpose of such transformation is to map 
the irregular integration space into a regular one, thereby avoid sample rejections for achieving higher efficiency.
The sampled points are passed to the evaluators for contributions by corresponding transformers. 
The calculated results are scaled with correct Jacobin, then are passed back to the \texttt{vegas} kernel. 

\begin{equation}
\label{eQpowder1}
S_{\text{vol}}(Q) = \int \mathrm{d}\omega
\int_0^{2\pi}  \mathrm{d} \phi 
\int_0^\pi \mathrm{d} \theta 
\int_{Q_{min}}^{Q_{max}} \mathrm{d} r  f(r, \theta, \phi) r^2 \sin\theta
\end{equation}

Here $f$ is the scattering intensity at the Q point that is defined in the spherical coordinate $(r, \theta, \phi)$.
The scattering function $S(Q |_{Q_{min}}^{Q_{max}},\omega)$ is accumulated as a histogram of $\omega$ alone side the integration.

Similarly, in the spherical surface mode, the integration is

\begin{equation}
\label{eQpowder2}
S_{\text{surf}}(Q) = \int \mathrm{d}\omega
\int_0^{2\pi}  \mathrm{d} \phi 
\int_0^\pi \mathrm{d} \theta 
 f(r, \theta, \phi) r^2 \sin\theta
\end{equation}

The surface integration is performed over a domain that has one less dimension than the space over which volume integration is performed. However, the surface integration is likely missing the gamma point where the small inelastic scattering is significant. It therefore should be specified very precisely by the users. By default, \texttt{BzScope} uses the volume integration scheme.

\subsubsection{Workers}
\label{ssWorkers}

Workers evaluate a given integrand and apply the adapted method in the Monte Carlo fashion. 
The ``run'' function of this type of class is launched by the Python build-in multiprocessing package.
Results and associated standard deviations are returned from the worker.
There are two workers in this version of code.

The \texttt{DebyeWallerWorker} is typically executed first to calculate the Debye-Waller exponents matrix. As a by-product, the PDOS for each atom is also analyzed.

Based on the calculated data, the \texttt{IsotropicAverageWorker} can initiate the evaluation of a specified scattering function at finite temperature. Furthermore, incoherent scattering functions beyond third order for individual atoms are generated in \texttt{NCrystal} from the vibrational density of states, and extended to several hundred orders until convergence of the total cross section is achieved.

\subsubsection{Workflow}

\begin{figure}[htbp]
    \centering
    \includegraphics[width=1\textwidth]{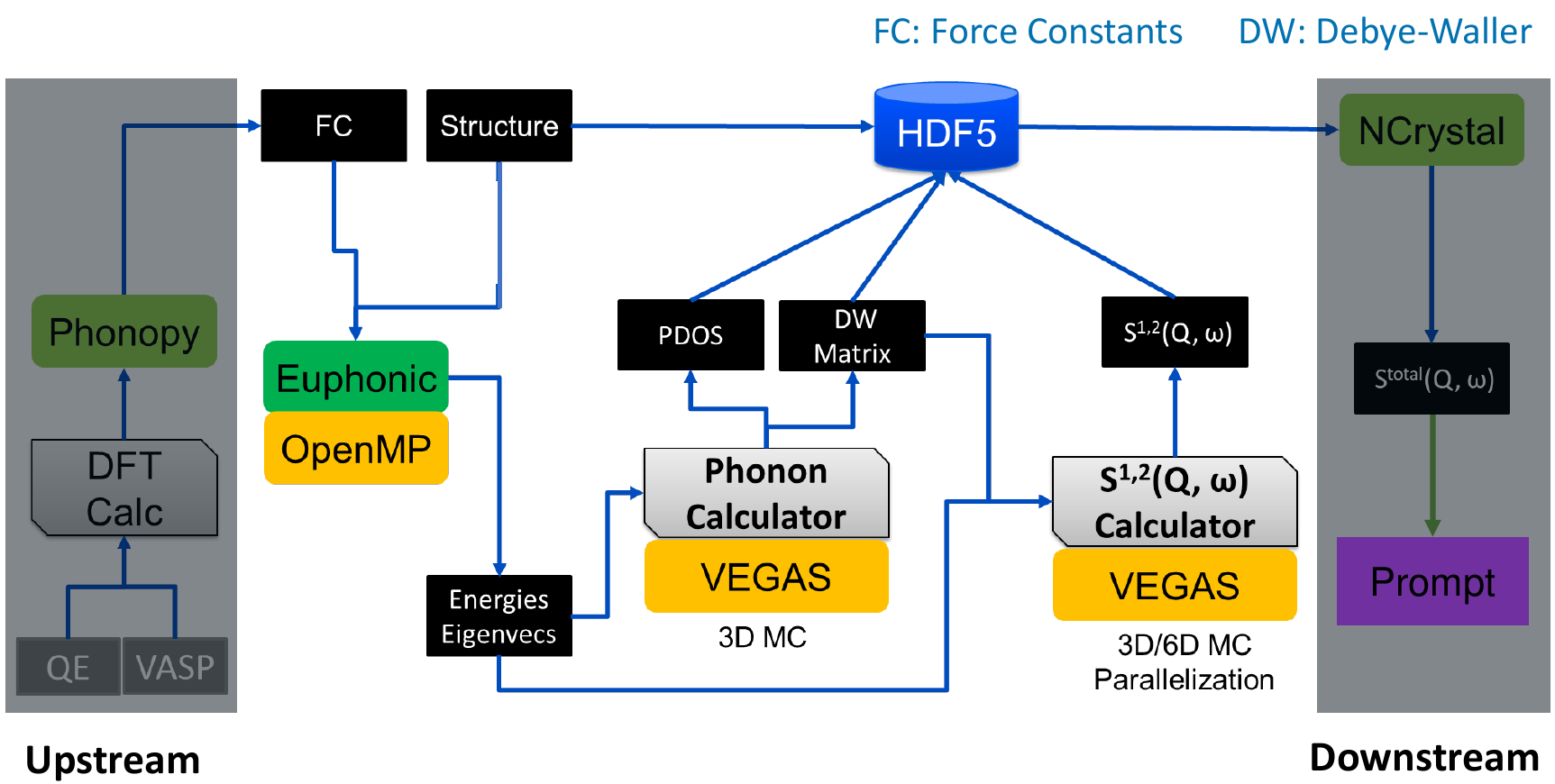}
    \caption{The workflow of \texttt{BzScope}.}
    \label{fig:workflow}
\end{figure}

The main workflow of \texttt{BzScope} is shown in Fig.~\ref{fig:workflow}.

Providing the force constants (FC) in the \texttt{phonopy} format and the structure information of the crystal sample, \texttt{BzScope} first calls \texttt{Euphonic} to parallelly obtain the energies and eigenvectors of phonons at given $q$ points or mesh. Then the \texttt{DebyeWallerWorker} runs to calculate PDOS and Debye-Waller exponents matrix, or the so-called DW matrix. 
Such a procedure is denoted phonon calculator in Fig.~\ref{fig:workflow}.
After that, the produced energies, eigenvectors and DW matrix are taken by $S(Q, \omega)$ calculator to obtain the  scattering functions of interest.
Currently, three kinds of scattering functions are supported, i.e. incoherent single phonon, coherent single phonon and coherent two phonon.
Finally, the structure information, the calculated PDOS, the calculated scattering function and associated standard deviation, together with the calculation configuration and some useful metadata, are saved into a \texttt{HDF5} file.

\texttt{BzScope} supports force constants in \texttt{phonopy} format by default. One can use various ab-initio tool-kits such as \texttt{VASP} and \texttt{Quantum-Espresso}(\texttt{QE}) through \texttt{Phonopy} to acquire the force constants.
For user's convenient,
a top-level tool, \textit{bz\_phonondb\_operator}  introduced in section~\ref{ssPhonondb}, can also directly use the force and structure data in online \texttt{PhononDB}~\cite{atztogo2023} datasets.

The calculated scattering functions in a \texttt{HDF5} formatted file can be further processed by another top-level top \textit{bz2nc} to obtain the complete inelastic scattering functions. 
The missing scattering contributions, incoherent and higher order coherent scatterings, in the \texttt{HDF5}  file are 
completed by the \texttt{NCrystal} calculation.
This tool generates a \texttt{NCrystal} data file that can be loaded in by any \texttt{NCrystal}-enabled Monte Carlo calculations through a plugin~\cite{bz-nc-plugin}.
Hence, it bridges the gap between experiment and calculation.

\subsubsection{Example python script}

\begin{figure}[htbp]
    \centering
    \includegraphics[width=1\textwidth]{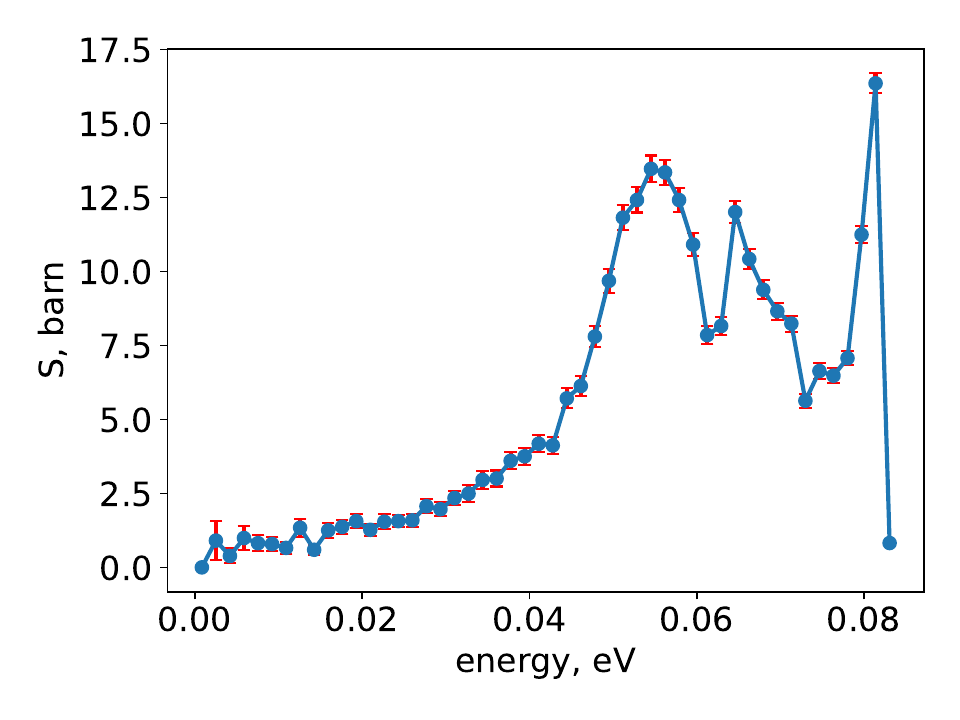}
    \caption{The result of the example script}
    \label{fig:example}
\end{figure}

The following example Python script in Listing~\ref{lst:pyexample} demonstrates the calculation of single phonon scattering cross section for Be at 30\si{\kelvin}, in the $Q$ range between 2.99 and 3.01 \si{\angstrom^{-1}}. 
The energy distribution in this $Q$ range is accumulated in 50 bins. 
The generated figure is shown in Fig.~\ref{fig:example}.

\begin{lstlisting}[caption={Sample Python}, label=lst:pyexample, language=Python]
#!/usr/bin/env python3

import matplotlib.pyplot as plt

from bzscope.calc.conf import CoherentPhononConf
from bzscope.calc.phonon import CoherentSinglePhonon
from bzscope.calc.integrand import PhononPowderAverageIntegrand
from bzscope.calc.worker import IsotropicAverageWorker

# It assumed that the crystal structure and force constants existed in directory Be.
calc_conf=CoherentPhononConf(temperature=30, min_Q=2.99,
                        max_Q=3.01, Q_bins=1, 
                        n_eval=1000, energy_bins=50,
                        phonopy_yaml_parent_path='./Be', 
                        phonopy_yaml_file='phonopy.yaml')
calc=CoherentSinglePhonon(calc_conf)
pwdIntegrand=PhononPowderAverageIntegrand(calc)
worker=IsotropicAverageWorker(pwdIntegrand, calc_conf.n_eval)
sqw, std=worker.run([0, False])

omega = calc.get_energy_centers()
plt.errorbar(omega, sqw, yerr=std, fmt='o-', 
             ecolor='r', capsize=3)
plt.xlabel(r'energy, eV')
plt.ylabel(r'S, barn')
plt.show()

\end{lstlisting}

\subsection{Top-level tools}

In addition to being a Python library, \texttt{BzScope} also offers several top-level tools that directly expose its full functionalities in the command line environment, as listed in Table~\ref{tab:top_level_tools}. These tools serve as \verb|Bash| or \verb|CMD| wrapper scripts for the corresponding Python modules and can alternatively be invoked using the syntax ``\verb|python3 -m bzscope.app.XXX|'', where ``\verb|XXX|'' denotes the tool name.

\begin{table}[htbp]
    \centering
    \caption{Top-level tools in \texttt{BzScope}}
    \label{tab:top_level_tools}
    \begin{tabular}{l p{7cm}}
        \toprule
        \multicolumn{1}{c}{\textbf{Name}} & \multicolumn{1}{c}{\textbf{Description}} \\
        \midrule
        \texttt{bz\_inelastic\_direct} &  This tool reads in the force constants in \texttt{phonopy} output format and outputs the PDOS and $S(Q, \omega)$ for powder samples. \\
        \texttt{bz\_inelastic\_indirect} & This tool reads in the output of \texttt{bz\_inelastic\_direct} and produces the 2D $S(Q, \omega)$ along a specified $(Q, \omega)$ path. \\
        \texttt{bz\_benchmark} & This tool is used to produces the benchmark results between \texttt{BzScope} and \texttt{NCrystal}. \\
        \texttt{bz\_phonondb\_operator} & This tool reads in the data from online \texttt{phonopy} \texttt{PhononDB} and outputs the force constants. \\
        \texttt{bz2nc} & This tool is used to produce ncmat file for \texttt{NCrystal} according to the PDOS calculated from \texttt{BzScope} \texttt{bz\_inelastic\_direct}. \\
        \bottomrule
    \end{tabular}
\end{table}

\subsubsection{bz\_inelastic\_direct}
This tool reads the phonopy formatted force constants and outputs PDOS and $S(Q, \omega)$, available parameters are shown in Table~\ref{tab:bz_direct}. Most of the options have default value set.

\begin{longtable}[c]{l p{4cm} l p{3cm}}
    \caption{Options available in \texttt{bz\_inelastic\_direct}.\label{tab:bz_direct}} \\
    \hline
    \multicolumn{1}{c}{\textbf{Name}} & \multicolumn{1}{c}{\textbf{Description}} &\multicolumn{1}{c}{\textbf{Unit}} & \multicolumn{1}{c}{\textbf{Default}} \\
    \hline
    \endfirsthead
    \multicolumn{4}{c}{Continuation of Table~\ref{tab:bz_direct}} \\
    \hline
    \multicolumn{1}{c}{\textbf{Name}} & \multicolumn{1}{c}{\textbf{Description}} &\multicolumn{1}{c}{\textbf{Unit}} & \multicolumn{1}{c}{\textbf{Default}} \\
    \hline
    \endhead
    \hline
    \endfoot
    \hline
    \endlastfoot
    \texttt{-h} & show help message and exit & - &  \\
    \texttt{-t TEMPERATURE} & temperature &  K &  \\
    \texttt{-c CUTOFF} & phonon cut off energy. Phonons with energy lower than this will be discarded. & eV & 0.0001 \\
    \texttt{-d DIRECTORY} & the directory from which the force constants and crystal structure loaded & string & . \\
    \texttt{-y PHONOPY\_YAML} & the phonopy yaml file & string & phonopy.yaml \\
    \texttt{-m MEASURED\_STRUCTURE} & the file from which the measured crystal structure is loaded & string & None \\
    \texttt{-l MIN\_Q} & lower limit for the Q & \AA\textsuperscript{-1} & 0.0 \\
    \texttt{-u MAX\_Q} & upper limit for the Q & \AA\textsuperscript{-1} & 10.0 \\
    \texttt{-q Q\_BINS} & Q bins for the histogram & int & 300 \\
    \texttt{-e ENERGY\_BINS} & energy bins for the histogram & int & 200 \\
    \texttt{-o OUTPUT} & output file name & string & bzdump.h5 \\
    \texttt{-n N\_EVAL} & number of evaluations per iteration & int & 10000 \\
    \texttt{-p PARTITIONS} & number of CPU cores & int & None (All cores) \\
    \texttt{-s} & integrate in the sphere surface mode & - & False \\
    \texttt{-x} & show the $S(Q, \omega)$ plot & - & False \\
    \texttt{--Qlog QLOG} & Q grid in log scale after this Q value & float & None \\
    \texttt{ --type TYPE} & calculation type, available options are c1, c2, i1, i1\_cubic & string & c1 \\
    \texttt{--use\_postfix} & use postfix in the output file name & - & False \\
\end{longtable}

\subsubsection{bz\_inelastic\_indirect}
This tool reads in the output file of \texttt{bz\_inelastic\_direct} and produces the 2D $S(Q, \omega)$ along a specified $(Q, \omega)$ path, available parameters are shown in Table~\ref{tab:bz_indirect}. Most of the options have default value set.

\begin{table}[htbp]
    \centering
    \caption{Options available in \texttt{bz\_inelastic\_indirect}}
    \label{tab:bz_indirect}
    \begin{tabular}{l p{6cm} l l}
        \toprule
        \multicolumn{1}{c}{\textbf{Name}} & \multicolumn{1}{c}{\textbf{Description}} &\multicolumn{1}{c}{\textbf{Unit}} & \multicolumn{1}{c}{\textbf{Default}} \\
        \midrule
        \texttt{-h} & show help message and exit & - &  \\
        \texttt{-e E\_OUT} & outcident neutron energy & eV & 0.00335 \\
        \texttt{-a ANGLE} & scattering angle & degree & 42.6 \\
        \texttt{-emin E\_IN\_MIN} & minimum incident neutron energy & eV & 0.00635 \\
        \texttt{-emax E\_IN\_MAX} & maximum incident neutron energy & eV & 0.50335 \\
        \texttt{-enum E\_IN\_NUM} & number of energy points & int & 1000 \\
        \texttt{-ai ATOM\_INDEX} & the index of atom to be calculated & int & None (all atoms) \\
        \texttt{-o OUTPUTFILE} & output file name & string & out \\
        \texttt{-i INSTRUMENT} & the instrument name, inline angle and energy will be used if provided. Currently only TFXA or TOSCA are supported. & string & None \\
        \texttt{-v} & show figures & - & False \\
        \texttt{-d} & debug mode & - & False \\
        \bottomrule
    \end{tabular}
\end{table}

\subsubsection{bz\_benchmark}
This tool is used to obtain the benchmark results between \texttt{BzScope} and \texttt{NCrystal}. It reads in the result file produced by \texttt{bz\_inelastic\_direct}. Available parameters are shown in Table~\ref{tab:bz_benchmark}.

\begin{table}[htbp]
    \centering
    \caption{Options available in \texttt{bz\_benchmark}}
    \label{tab:bz_benchmark}
    \begin{tabular}{l l l l}
        \toprule
        \multicolumn{1}{c}{\textbf{Name}} & \multicolumn{1}{c}{\textbf{Description}} &\multicolumn{1}{c}{\textbf{Unit}} & \multicolumn{1}{c}{\textbf{Default}} \\
        \midrule
        \texttt{-h} & show help message and exit & - &  \\
        \texttt{-c QCUT [...] } & Cut off value for the Q plots & \AA\textsuperscript{-1} & [80.0] \\
        \texttt{-o OUTPUTFILE} & the output ncmat file name & String & bztool \\
        \texttt{-v} & show figures & - & False \\
        \bottomrule
    \end{tabular}
\end{table}

\subsubsection{bz\_phonondb\_operator}
\label{ssPhonondb}
This tool retrieves chosen datasets from the online \texttt{phonopy} \texttt{PhononDB} according to the dataset identifier and includes the force constants to the output file. The output can then be used by \texttt{bz\_inelastic\_direct}. Available parameters are shown in Table~\ref{tab:bz_phonondb}.

\begin{table}[htbp]
    \centering
    \caption{Sub-commands  and options available in \texttt{bz\_phonondb\_operator}}
    \label{tab:bz_phonondb}
    \begin{tabular}{l l p{5cm} l l}
        \toprule
        \multicolumn{1}{c}{\textbf{Sub-command}} & \multicolumn{1}{c}{\textbf{Name}} & \multicolumn{1}{c}{\textbf{Description}} & \multicolumn{1}{c}{\textbf{Unit}} & \multicolumn{1}{c}{\textbf{Default}} \\
        \midrule
        s & & retrieve a single dataset & & \\
         & \texttt{-h} & show help message and exit & - &  \\
         & \texttt{-u} & force to download the dataset from PhononDB, regardless the cache & - & False \\
         & \texttt{-n} & do not calculate the force constants & - & False \\
        \midrule
        t & & retrieve all datasets & & \\
         & \texttt{-h} & show help message and exit & - &  \\
         & \texttt{-u} & force to download the dataset from PhononDB, regardless the cache & - & False \\
         & \texttt{-c} & calculate the force constants after download & - & False \\
        \bottomrule
    \end{tabular}
\end{table}

\subsubsection{bz2nc}
This tool is used to produce ncmat file for using by \texttt{NCrystal} according to the PDOS calculated from \texttt{BzScope} \texttt{bz\_inelastic\_direct}. Available parameters are shown in Table~\ref{tab:bz_bz2nc}.

\begin{table}[htbp]
    \centering
    \caption{Options available in \texttt{bz2nc}}
    \label{tab:bz_bz2nc}
    \begin{tabular}{l l l l}
        \toprule
        \multicolumn{1}{c}{\textbf{name}} & \multicolumn{1}{c}{\textbf{description}} &\multicolumn{1}{c}{\textbf{unit}} & \multicolumn{1}{c}{\textbf{default}} \\
        \midrule
        \texttt{-h} & show help message and exit & - &  \\
        \texttt{-o OUTPUTFILE} & output ncmat file name & string &  \\
        \texttt{-v} & show figures & - & False \\
        \bottomrule
    \end{tabular}
\end{table}

\subsubsection{A simple example}
The bash script in Listing~\ref{lst:example_bash} demonstrates procedures of downloading phonon data of Silicon (mp-149) from the \texttt{PhononDB} and calculating the scattering function and benchmark.  

\begin{lstlisting}[language=bash, caption={Sample Bash}, label=lst:example_bash]
#!/bin/bash
set -e
# First download the dataset from PhononDB repository
# and calculate the force constants, which will produce
# the phonopy.yaml file in sub-directory m039k924j
python3 -m bzscope.app.bz_phonondb_operator s m039k924j
# Then calculate the scattering function for c1, which 
# will produce bzdump.h5
python3 -m bzscope.app.bz_inelastic_direct -d m039k924j -t 300 -u 40 -x
# (Optional) Then calculate the 2D sqw along a specified 
# path by loading bzdump.h5. In this example, TOSCA is 
# used as the instrument. Two files will be produced:
# out.pdf and out_merged_sqw.h5
python3 -m bzscope.app.bz_inelastic_indirect -i tosca -v
# Then do the benchmark, which will read in bzdump.h5
# and produce bztool.pdf
python3 -m bzscope.app.bz_benchmark -c 40 -v
# Finally generate the ncmat file for NCrystal and show
# plots
python3 -m bzscope.app.bz2nc -v -o si149
\end{lstlisting}

\section{Benchmark}

In order to get a comprehensive evaluation of the numerical performance of BzScope, materials belonging to different crystal systems and space groups (SG) are carefully chosen, which vary from cubic to monoclinic structures. Computed results at different temperatures are also compared with existing tools or experiments.
The numerical performance of \texttt{BzScope} is benchmarked against the robust incoherent approximation model in NCrystal in sections~\ref{ssINcoh} to~\ref{ssCOherent2}.
Calculated differential and total sections are benchmarked against experimental data in section~\ref{ssDiffXS} and~\ref{ssTotXS}. 
 
In this section, six initial crystal structures from Materials Project~\cite{atsushi-lih} (with IDs represented as mp-XXX) are used, as summarised in Table~\ref{tab:benchmark}.

For nickel (Ni) and beryllium (Be),  Phonopy v2.15.0 is used to make the input script for Quantum-Espresso (QE) v6.4.1 by the finite displacement method. 
This is followed by the relaxation calculation and the self-consistent field calculation carried out by QE to get the relaxed structure and forces. The precision category of standard solid-state pseudopotentials (SSSP)~\cite{prandini2018precision} is used in the  calculation. 

For silicon (Si), nickel phosphide (NiP2) and lithium hydride (LiH), the crystal structures and forces are obtained directly from the online Phonopy PhononDB~\cite{atztogo2023}, of which the initial structures are also obtained from the Materials Project, generated by Phonopy. 

\begin{table}[htbp]
    \centering
    \caption{\texttt{BzScope} calculated crystals for this section. Notice that for the force constant that calculated by finite displacement method of Phonopy, QE is used for the DFT calculations.  }
    \begin{tabular}{l c c c r}
        \toprule
        \multicolumn{1}{c}{\textbf{Material}} & 
        \multicolumn{1}{c}{\textbf{Space group}} & 
        \multicolumn{1}{c}{\textbf{MP entry}} &
        \multicolumn{1}{c}{\textbf{Source}} &\multicolumn{1}{c}{\textbf{Temperature}} \\
        &&&& (K) \\
        \midrule
        \texttt{Ni} & cubic (225) & mp-23 & QE + Phonopy\tablefootnote{The supercell and k-mesh sizes are respectively $4\times4\times4$ and $3\times3\times3$.} &20, 77, 300 \\
        \texttt{Be} & hexagonal (194) & mp-87  & QE + Phonopy\tablefootnote{The supercell and k-mesh sizes are respectively $4\times4\times2$ and $3\times3\times4$.}& 77, 300 \\
        \texttt{Si} & cubic (227) & mp-149& PhononDB m039k924j &  300 \\
        \texttt{Si} & hexagonal (194) & mp-165 & PhononDB k930c3397 & 77 \\
        \texttt{\(\text{NiP}_2 \)} & mono-clinic (15) & mp-486 &  PhononDB  tb09jb156 & 77 \\
        \texttt{LiH} & cubic (225) & mp-23703 & PhononDB tt44ps026 & 20 \\
        \bottomrule
    \end{tabular}
    \label{tab:benchmark}
\end{table}

\subsection{Self-adaptive sampling the vibrational density of states}

\begin{figure}[htbp]
  \centering
  \includegraphics[width=0.6\textwidth]{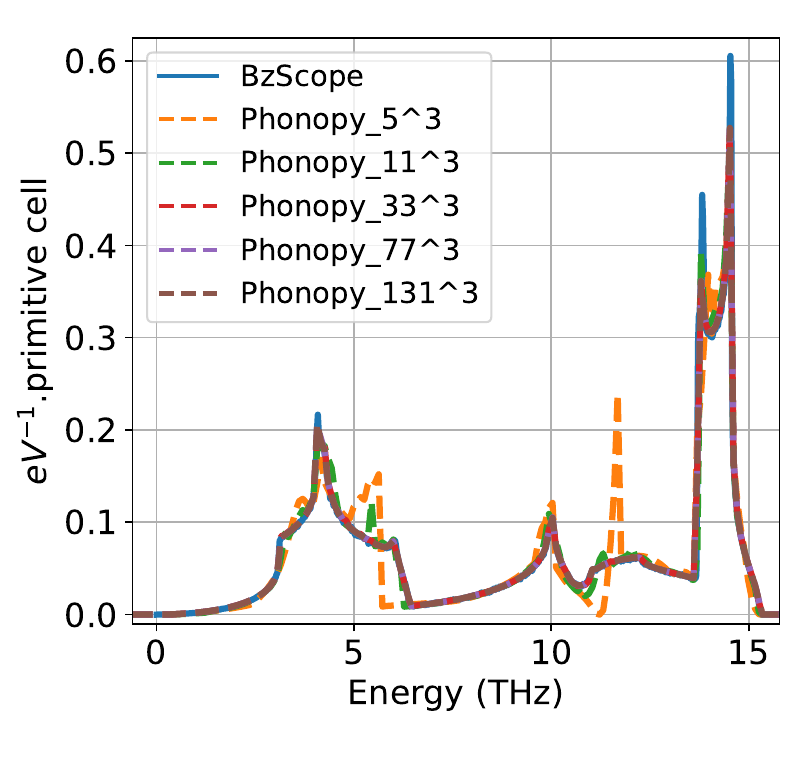}
  \caption{Calculated VDOS of cubic Si (mp-149) at \SI{300}{\kelvin}}
  \label{fig:si_vdos}
\end{figure}

To test the self-adaptive Monte Carlo sampling ability of \texttt{BzScope}, the VDOS of cubic silicon is calculated in comparison with the regular grid method shown in Fig.~\ref{fig:si_vdos}. \texttt{BzScope} can produce the optimal VDOS in just one go, hence freeing users from convergence tests for different mesh sizes. 

\subsection{Incoherent single-phonon scattering function}
\label{ssINcoh}

\texttt{BzScope} is used to calculate the  incoherent single phonon scattering function for face-centered cubic nickel, treated as a six-dimensional problem, in which both the $\bm Q$ and $\bm q$ points are sampled independently. As it is a type of cubic Bravais lattice, the incoherent model of NCrystal outlined in section~\ref{ssCubicIncoAppr} is expected to be strictly equivalent to the calculation of BzScope. 
Hence these comparisons show the numerical performance of BzScope.

The scattering function is calculated in the Q range from 0 to \SI{80}{\per\angstrom} linear spaced in 800 bins. 
For each bin, the scattering function is evaluated 10000 times in each of the 10 integration batches.
The integral of $ \int^{Q_{+}}_{Q_-} S(\bm Q, \omega) \, d\omega $ is used as the target of the integration. 
During the Monte Carlo calculation in a given Q range, the energy distribution is estimated in a 400-bin histogram.

\begin{figure}[htbp]
  \centering
  \begin{subfigure}[b]{0.45\textwidth}
    \includegraphics[width=\textwidth, page=1]{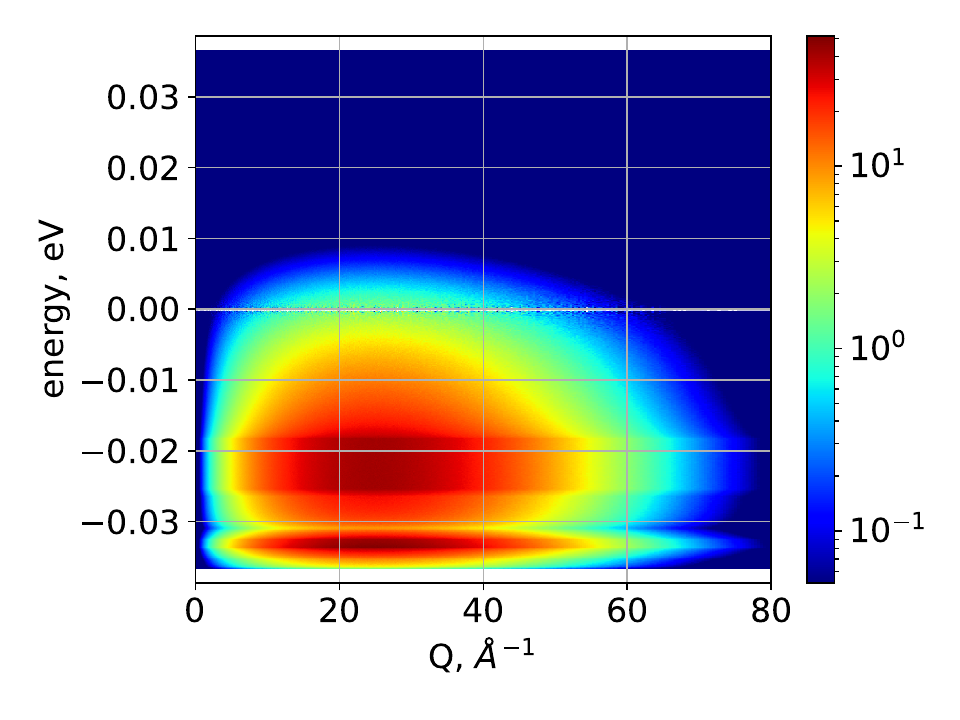}
    \caption{BzScope}
    \label{fig:ni_i1_20K_a}
  \end{subfigure}
  \hfill
  \begin{subfigure}[b]{0.45\textwidth}
    \includegraphics[width=\textwidth, page=2]{img/ni_i1_20K.pdf}
    \caption{NCrystal}
    \label{fig:ni_i1_20K_b}
  \end{subfigure}
  \caption{Incoherent single-phonon scattering function of Ni at 20K}
  \label{fig:ni_i1_20K}
\end{figure}

Fig.~\ref{fig:ni_i1_20K_a} and~\ref{fig:ni_i1_20K_b} show the single phonon incoherent scattering function of Ni calculated by \texttt{BzScope} and NCrystal, respectively. 
The scattering functions seem to be remarkable similar, if not identical, except that the intensity in the \texttt{BzScope} result is almost zero when energy transfer approaches zero. Such an observation is caused by the $1/\omega$  factor when evaluating Eq.~\ref{eIC1}. For numerical stability, contributions from phonons that are less than 
 \SI{0.1}{\milli\electronvolt} are discarded in \texttt{BzScope}.
 While in NCrystal, Eq.~\ref{eHnFunc} is Taylor expanded when energy is close to zero. By assuming the power law shape of the density of states close to the gamma points, specifically $\rho \propto \omega^2$ in the Debye model, NCrystal is able to cancel out the $1/\omega$ factor when energy transfer is approaching zero.
 
\begin{figure}[htbp]
  \centering
  \begin{subfigure}[b]{0.45\textwidth}
    \includegraphics[width=\textwidth, page=3]{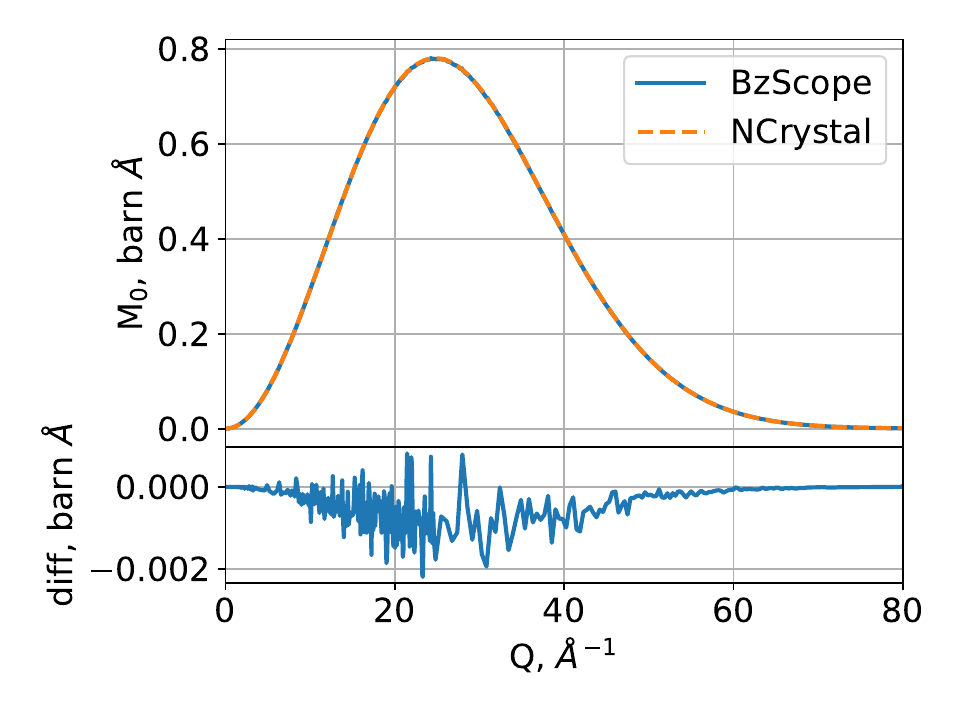}
    \caption{20K}
    \label{fig:ni_i1_m0_20K_a}
  \end{subfigure}
  \hfill
  \begin{subfigure}[b]{0.45\textwidth}
    \includegraphics[width=\textwidth, page=3]{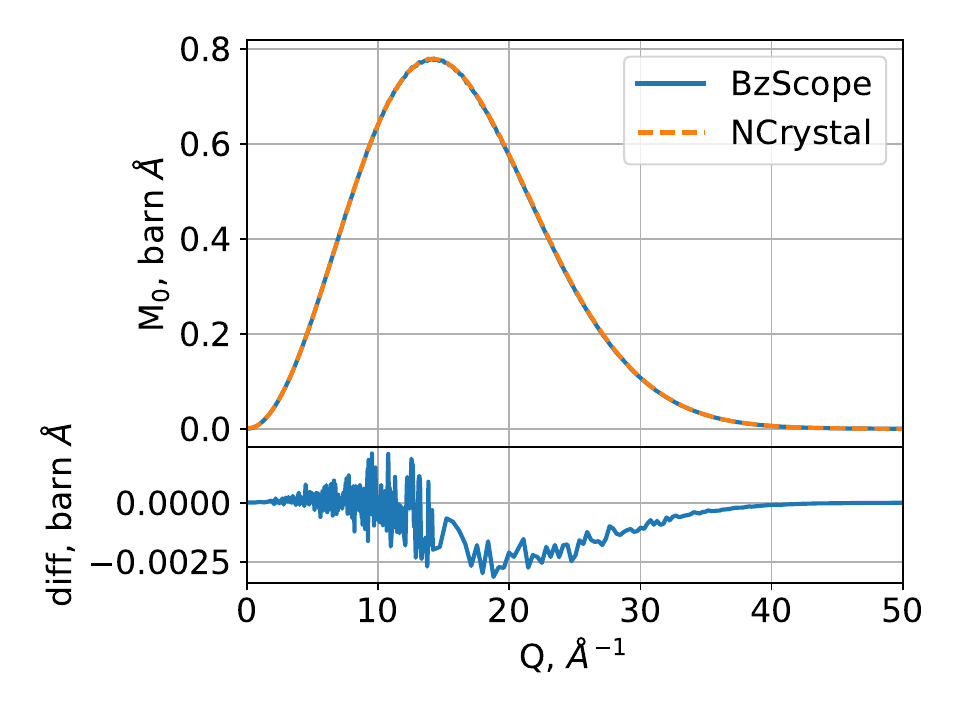}
    \caption{300K}
    \label{fig:ni_i1_m0_300K_b}
  \end{subfigure}
  \caption{Zeroth moment $M_0(Q)$ of the incoherent single-phonon scattering function in  Ni at different temperature}
  \label{fig:ni_i1_Zeroth}
\end{figure}

The $n$-th order moment function for a  partial scattering function is defined as $M_n(Q)=\int_{-\infty}^{+\infty} (\hbar \omega)^n S_p(Q, \omega) d\,\omega$, where $S_p$ is for a particular coherent or incoherent scattering function.
The zeroth order momentum functions for Ni in Fig.~\ref{fig:ni_i1_20K} are compared in  Fig~\ref{fig:ni_i1_Zeroth}, for the better visualization purpose. 
The discrepancy is better than 0.3\% in the regions that have the most contributions at both \SI{20} and \SI{300}{\kelvin}.
However, \texttt{BzScope} almost always slightly underestimates the intensity, indicating that a bias exists in the calculation. 
Such observation is the result of the mentioned  numerical treatment for phonons close to the gamma point, where their energies are insignificant.
 
Similarly, the incoherent $M_0(Q)$  curves of two more crystals, i.e. Be and  \( \text{NiP}_2 \), from the less symmetric space groups are evaluated to put the isotropic displacement approximation in NCrystal in the test. 

\begin{figure}[htbp]
  \centering
  \begin{subfigure}[b]{0.45\textwidth}
    \includegraphics[width=\textwidth, page=3]{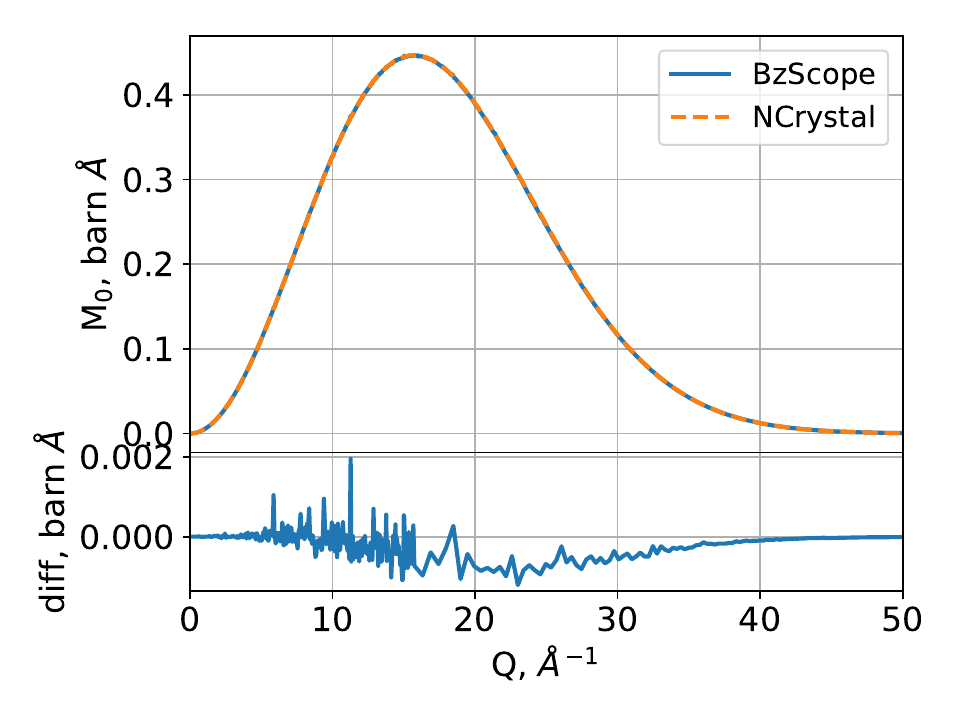}
    \caption{Be}
    \label{fig:be_i1_77K_a}
  \end{subfigure}
  \hfill
  \begin{subfigure}[b]{0.45\textwidth}
    \includegraphics[width=\textwidth, page=3]{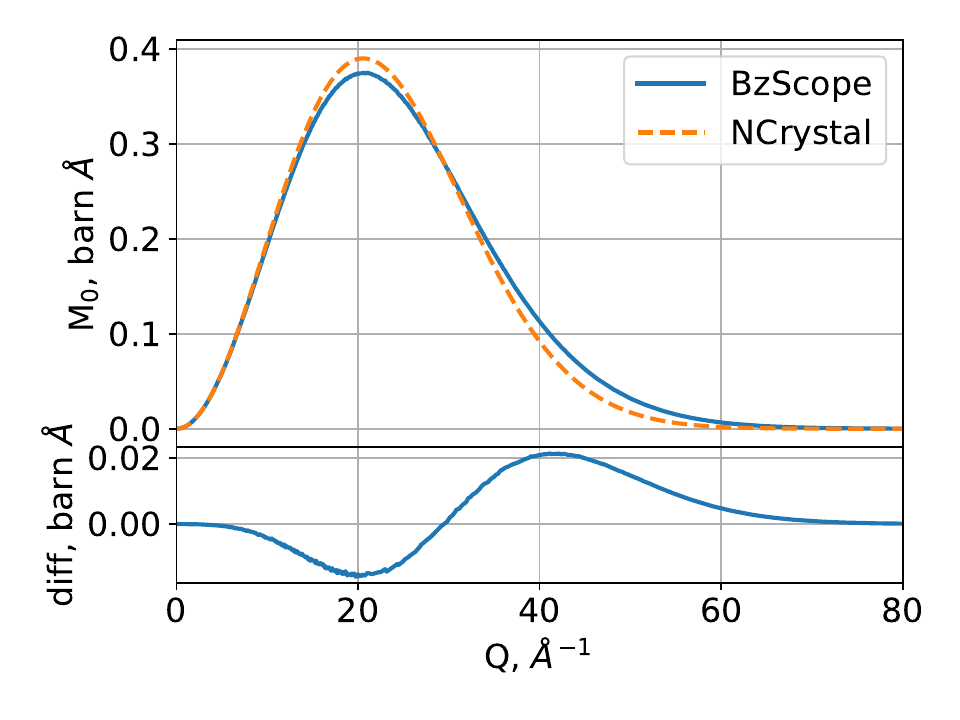}
    \caption{\( \text{NiP}_2 \)}
    \label{fig:nip2_i1_77K_b}
  \end{subfigure}
  \caption{Zeroth moment $M_0(Q)$ of the incoherent single-phonon scattering function in Be of space group P6\_3$/$mmc(194) and \( \text{NiP}_2 \) of space group C2$/$c(15). Both at \SI{77}{\kelvin}.  Compare with the results from NCrystal incoherent model.}
  \label{fig:be_nip2_77k}
\end{figure}

Fig.~\ref{fig:be_nip2_77k} shows the zeroth moment for Be and \( \text{NiP}_2 \). 
In Fig.~\ref{fig:be_i1_77K_a}, the approximation holds quit well even for this hexagonal close-packed crystal. The discrepancies between NCrystal and \texttt{BzScope} are less than 1\%, having similar behaviour as the Ni cases.

However, in Fig.~\ref{fig:nip2_i1_77K_b}, the incoherent case  for  \( \text{NiP}_2 \), the approximation in NCrystal is no longer held.
Even though the results are broadly similar, both models predict a broad peak situated around \SI{20}{\per\angstrom}.
The differential curve is composed of a wide dip that occurs below approximately \SI{20}{\per\angstrom}, along with a wide peak that appears above this value at about \SI{43}{\per\angstrom}.
The observation is mainly due to the mismatch between the isotropic displacements assumption used in the \text{NCrystal} calculations and the actual low-symmetry material. The difference can be as great as 20\%. The telegraph like, but negligible, numerical discrepancies still appear, noticeably at around \SI{20}{\per\angstrom}.

\subsection{Coherent single-phonon scattering function}
\label{ssCOherent}

\begin{figure}[htbp]
    \centering
    \begin{subfigure}[b]{0.45\textwidth}
    \includegraphics[width=\textwidth, page=3]{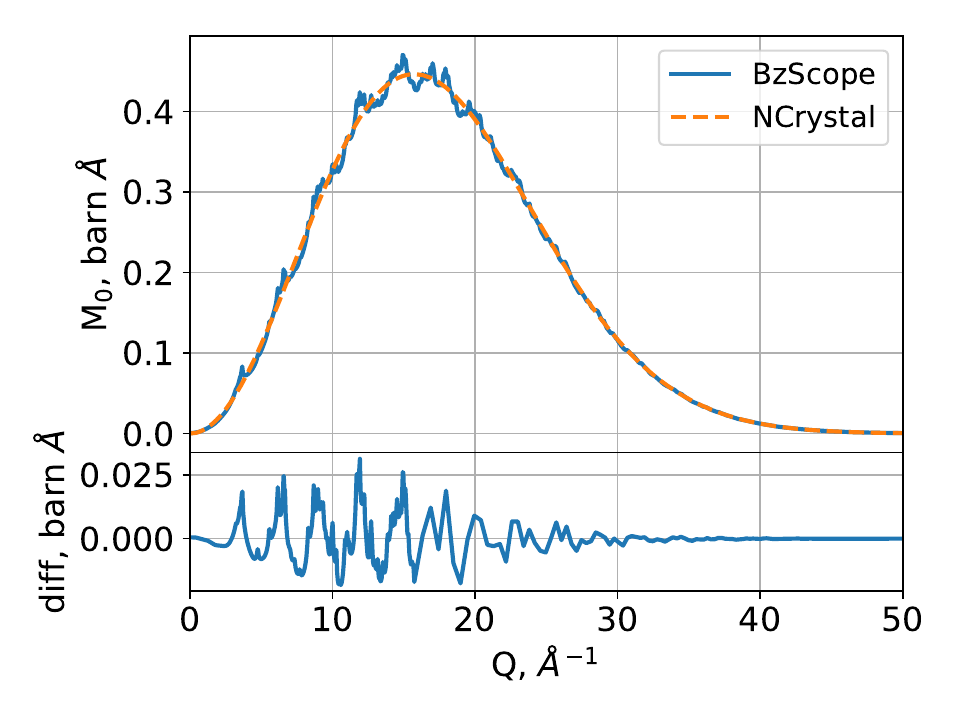}
    \caption{Be}
    \label{fig:be_c1_77K_a}
    \end{subfigure}
    \hfill
    \begin{subfigure}[b]{0.45\textwidth}
    \includegraphics[width=\textwidth, page=3]{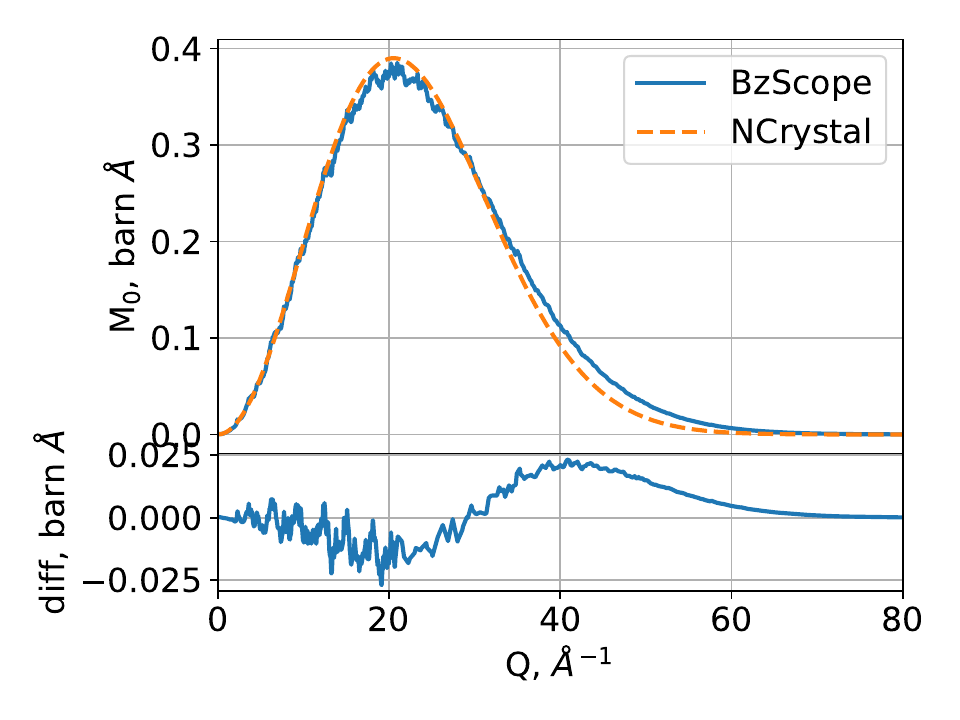}
    \caption{ \( \text{NiP}_2 \)}
    \label{fig:nip2_c1_77K_b}
    \end{subfigure}
    \caption{Zeroth moment  $M_0(Q)$ of the coherent single-phonon scattering in Be and \(\text{NiP}_2 \). Both at \SI{77}{\kelvin}. Compare with the results from NCrystal incoherent model.}
    \label{fig:be_nip2_c1_zeroth}
\end{figure}

The zeroth moments of the coherent single phonon scattering function of the two crystals, Be and  \( \text{NiP}_2 \), are also calculated, as shown in Fig.~\ref{fig:be_nip2_c1_zeroth}.

The \texttt{BzScope} results are broadly similar to the results from those calculated by NCrystal incoherent model, while with the additional fast varying distinct static structure factor. The discrepancies between \texttt{BzScope} and NCrystal are up to about 5\% and 25\% for Be and \( \text{NiP}_2 \), correspondingly.
Notice that the curves calculated by \texttt{BzScope} have many obvious peaks.
It is expected that the numerical accuracy is similar to that in the incoherent case.
The observed peaks are not numerical errors. 

\begin{figure}[htbp]
    \centering
    \includegraphics[width=\textwidth]{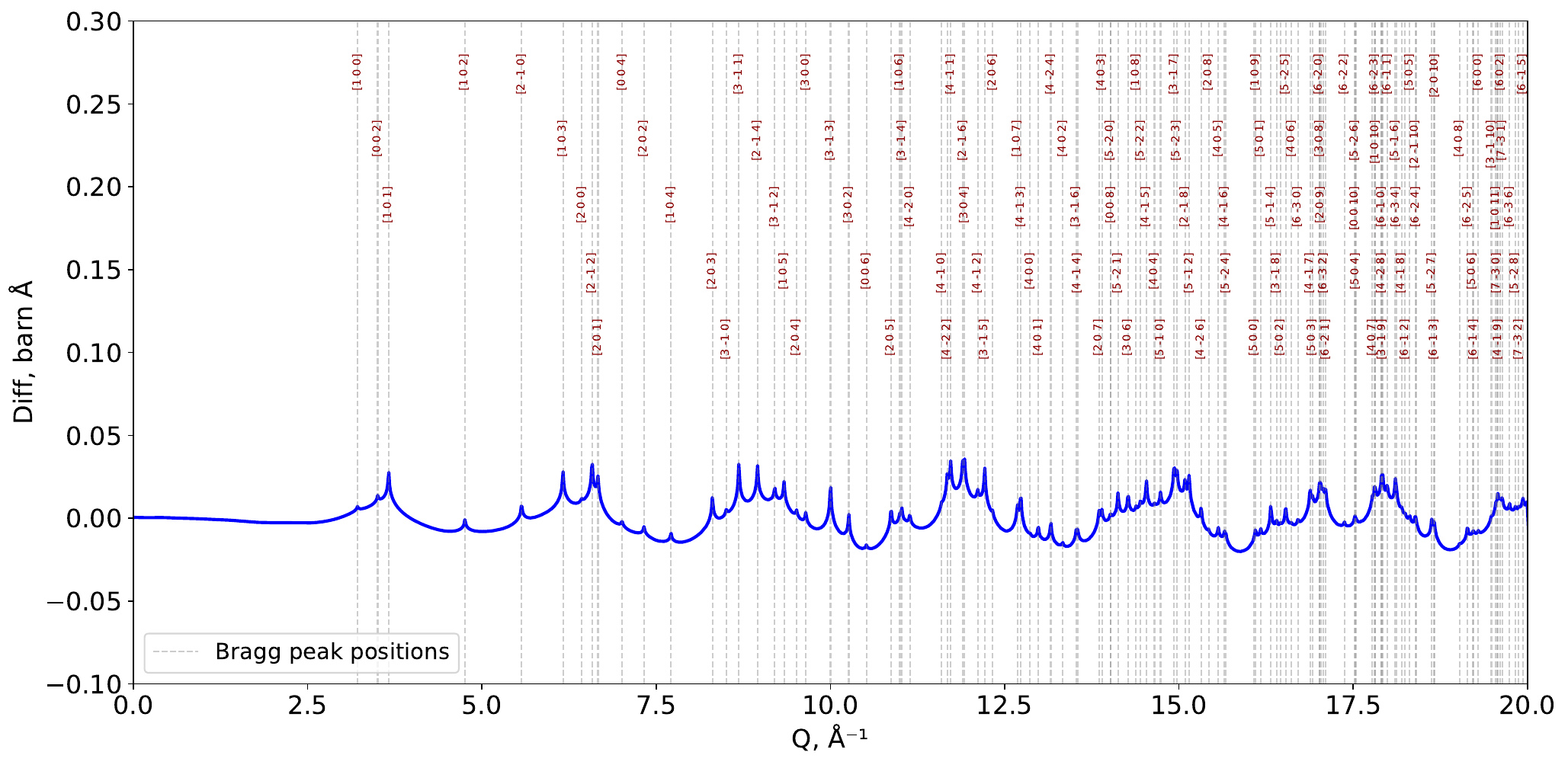}
    \caption{$M_0$ difference of \text{Be} with the Bragg peak positions}
    \label{fig:be_diff_analysis}
\end{figure}

Shown with higher resolution, the $M_0$ difference between the coherent and incoherent approximated single phonon scattering function in Fig.~\ref{fig:be_c1_77K_a} is shown along with the Bragg peak positions that predicted by NCrystal in Fig.~\ref{fig:be_diff_analysis}. The upper limit of $Q$ is set to \SI{20}{\angstrom^{-1}} due to the increasing peak density at higher $Q$, which makes individual peaks difficult to resolve.
The structures in the difference are systematically aligned with the positions, indicating that the fluctuations are primarily caused by the coherent single phonon scattering processes.
Therefore, the peaks are resulting from  the distinct correlation function, indicating that \texttt{BzScope} has higher accuracy after incorporating the contribution of coherent scattering.

\subsection{Coherent two-phonon scattering function}
\label{ssCOherent2}

The comparison between results from \texttt{BzScope} and \texttt{NCrystal} for coherent two phonons scattering function is also performed.

\begin{figure}[htbp]
  \centering
  \begin{subfigure}[b]{0.45\textwidth}
    \includegraphics[width=\textwidth, page=1]{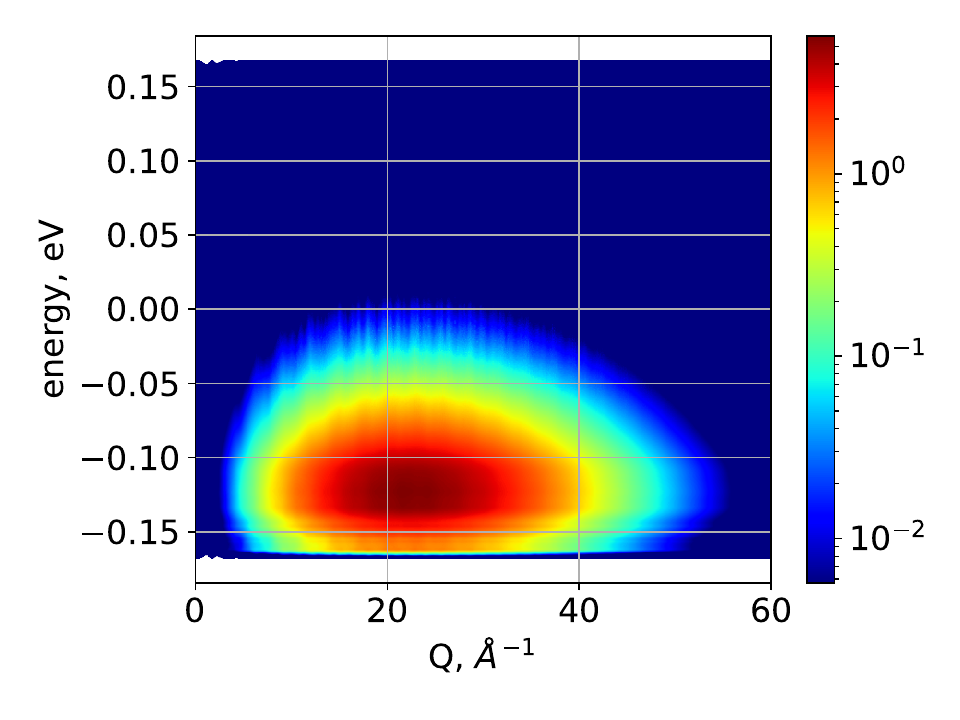}
    \caption{\(\texttt{BzScope}\)}
    \label{fig:be_c2_77k_map_a}
  \end{subfigure}
  \hfill
  \begin{subfigure}[b]{0.45\textwidth}
    \includegraphics[width=\textwidth, page=2]{img/be_c2_77k.pdf}
    \caption{\(\texttt{NCrystal}\)}
    \label{fig:be_c2_77k_map_b}
  \end{subfigure}
  \caption{Intensity map of the coherent double-phonon scattering function in \(\text{Be}\) at \SI{77}{\kelvin}. Compare with the results from NCrystal incoherent model.}
  \label{fig:be_c2_77k_map}
\end{figure}

Fig.~\ref{fig:be_c2_77k_map} shows the two phonon coherent scattering function $S(Q, \omega)$ from these two packages. Notice that such function is approximated by the incoherent model by NCrystal. Subsequently, the scattering intensity along the momentum transfers is smoother. 
By contrast, \texttt{BzScope} calculated scattering function shows detailed but weak structures along that dimension, likely originated from the distinct pair correlation.    

Fig.~\ref{fig:be_nip2_c2_zeroth} shows the {zeroth moment  $M_0(Q)$ of the coherent double phonons scattering function for both \(\text{Be}\) and \(\text{NiP}_2\). The results from both models agree well with each other for \(\text{Be}\) (high symmetry), with the discrepancy far better than 1\%. As the peak positions in the difference are not correlated with the positions of the Bragg peaks, these small structure are likely the numerical error.  
While for \(\text{NiP}_2\), the discrepancy goes up to 25\% due to the isotropic displacement approximation failed in this low symmetry crystal.  The numerical error is insignificant in comparison with the discrepancies introduced by physics.

\begin{figure}[htbp]
  \centering
  \begin{subfigure}[b]{0.45\textwidth}
    \includegraphics[width=\textwidth, page=3]{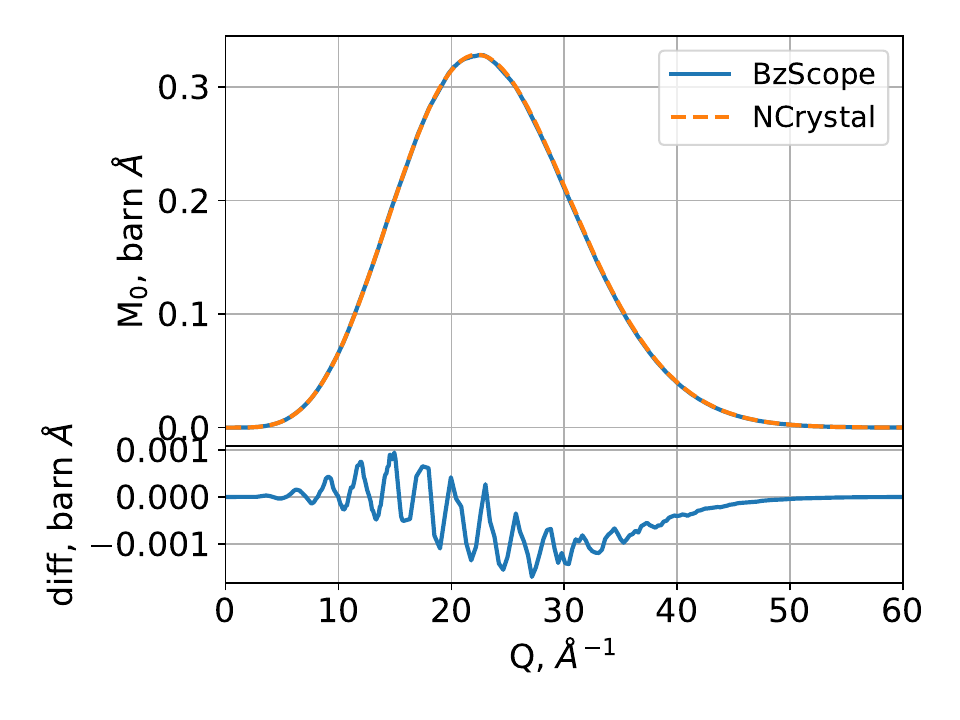}
    \caption{\(\text{Be} \)}
    \label{fig:be_nip2_c2_77k_c}
  \end{subfigure}
  \hfill
  \begin{subfigure}[b]{0.45\textwidth}
    \includegraphics[width=\textwidth, page=3]{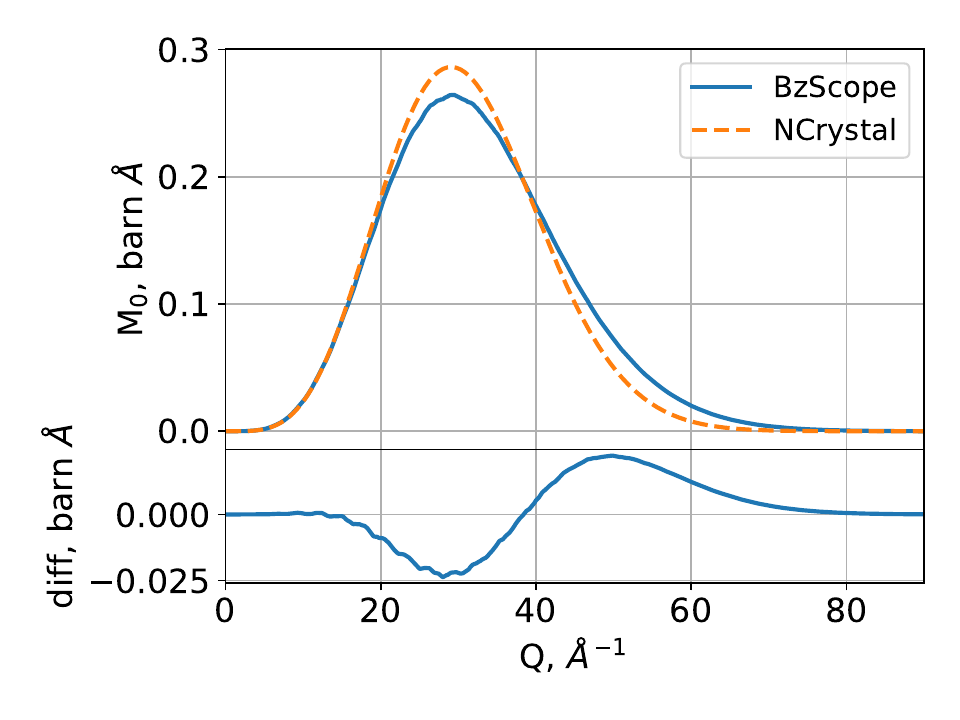}
    \caption{\(\text{NiP}_2 \)}
    \label{fig:be_nip2_c2_77k_d}
  \end{subfigure}
  \caption{Zeroth moment  $M_0(Q)$ of the coherent double-phonon scattering function in \(\text{Be}\) and \(\text{NiP}_2\). Both at \SI{77}{\kelvin}. Compare with the results from NCrystal incoherent model.}
  \label{fig:be_nip2_c2_zeroth}
\end{figure}

Fig.~\ref{fig:ni_c2_zeroth} shows the {zeroth moment  $M_0(Q)$ of the coherent double-phonon scattering function in \(\text{Ni}\) at 20, 77 and 300\si{\kelvin}. 
The maximum discrepancy clearly grows with temperature,  from about 0.5\% at 20\si{\kelvin} to about 5\% at 300\si{\kelvin}.
As observed that the numerical error in the benchmark setup is in the sub-percentage range, the main source of discrepancies at higher temperature is likely the distinct pair correlation.

Zoom-in the difference of Fig.~\ref{fig:ni_c2_c} and place the Bragg diffraction positions in Fig.~\ref{fig:ni_300k_diff_analysis}.
It can be observed that the peak positions are not precisely correlated with each Bragg diffraction. In comparison with the single phonon scattering case in Fig.~\ref{fig:be_diff_analysis}, the structures are significantly broader in the two phonon scattering case. 
This implies that the shape structure for single phonon scattering are broadened in the convolution process of two phonon scattering.

\begin{figure}[htbp]
    \centering
    \begin{subfigure}[b]{0.32\textwidth}
    \includegraphics[width=\textwidth, page=3]{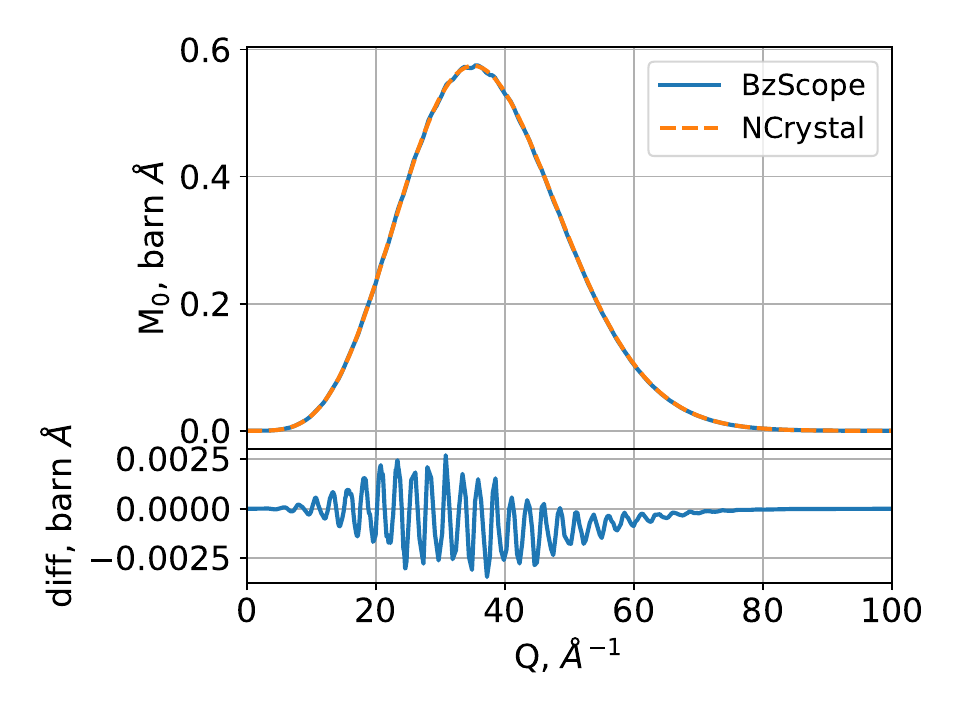}
    \caption{\SI{20}{\kelvin}}
    \label{fig:ni_c2_a}
    \end{subfigure}
    \hfill
    \begin{subfigure}[b]{0.32\textwidth}
    \includegraphics[width=\textwidth, page=3]{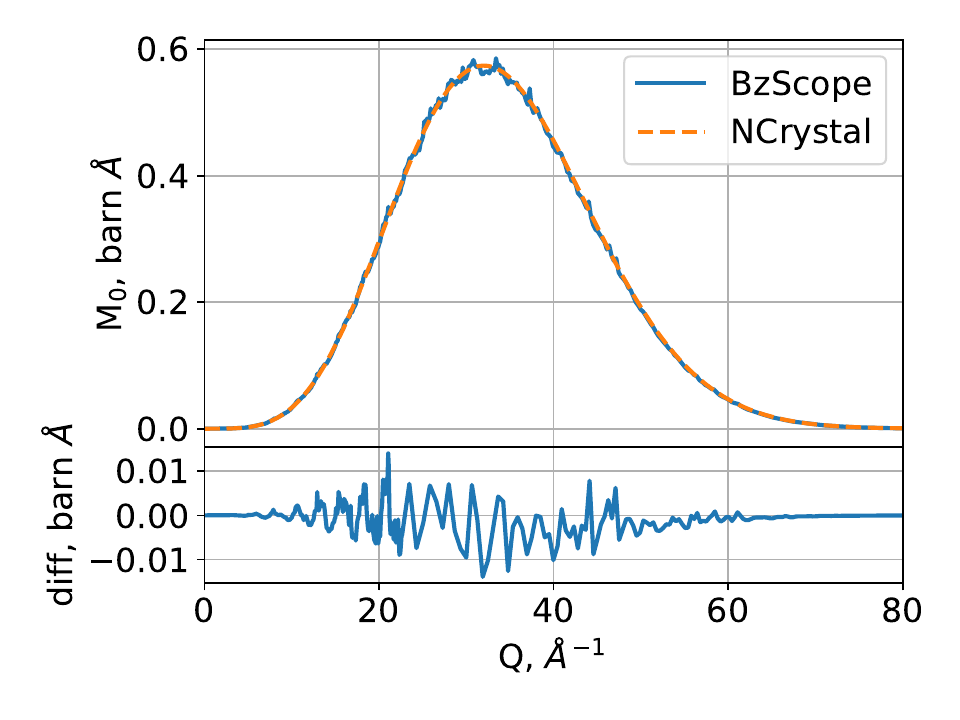}
    \caption{\SI{77}{\kelvin}}
    \label{fig:ni_c2_b}
    \end{subfigure}
    \hfill
    \begin{subfigure}[b]{0.32\textwidth}
    \includegraphics[width=\textwidth, page=3]{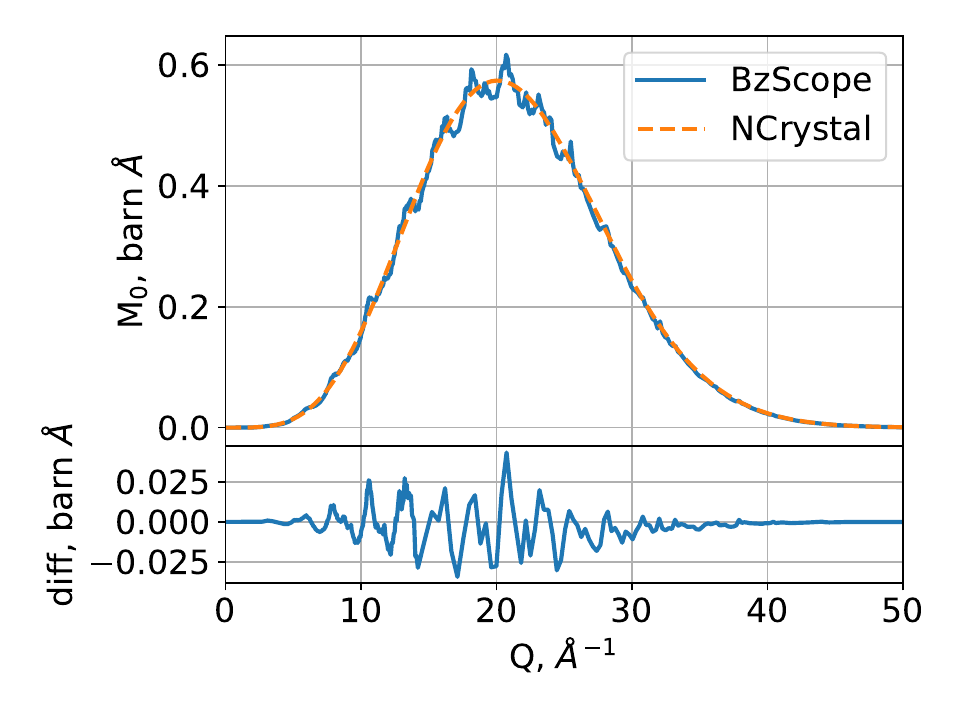}
    \caption{\SI{300}{\kelvin}}
    \label{fig:ni_c2_c}
    \end{subfigure}
    \caption{Zeroth moment  $M_0(Q)$ of the coherent double-phonon scattering function in \(\text{Ni}\) at different temperature. Compare with the results from NCrystal incoherent model.}
    \label{fig:ni_c2_zeroth}
\end{figure}

\begin{figure}[htbp]
    \centering
    \includegraphics[width=\textwidth]{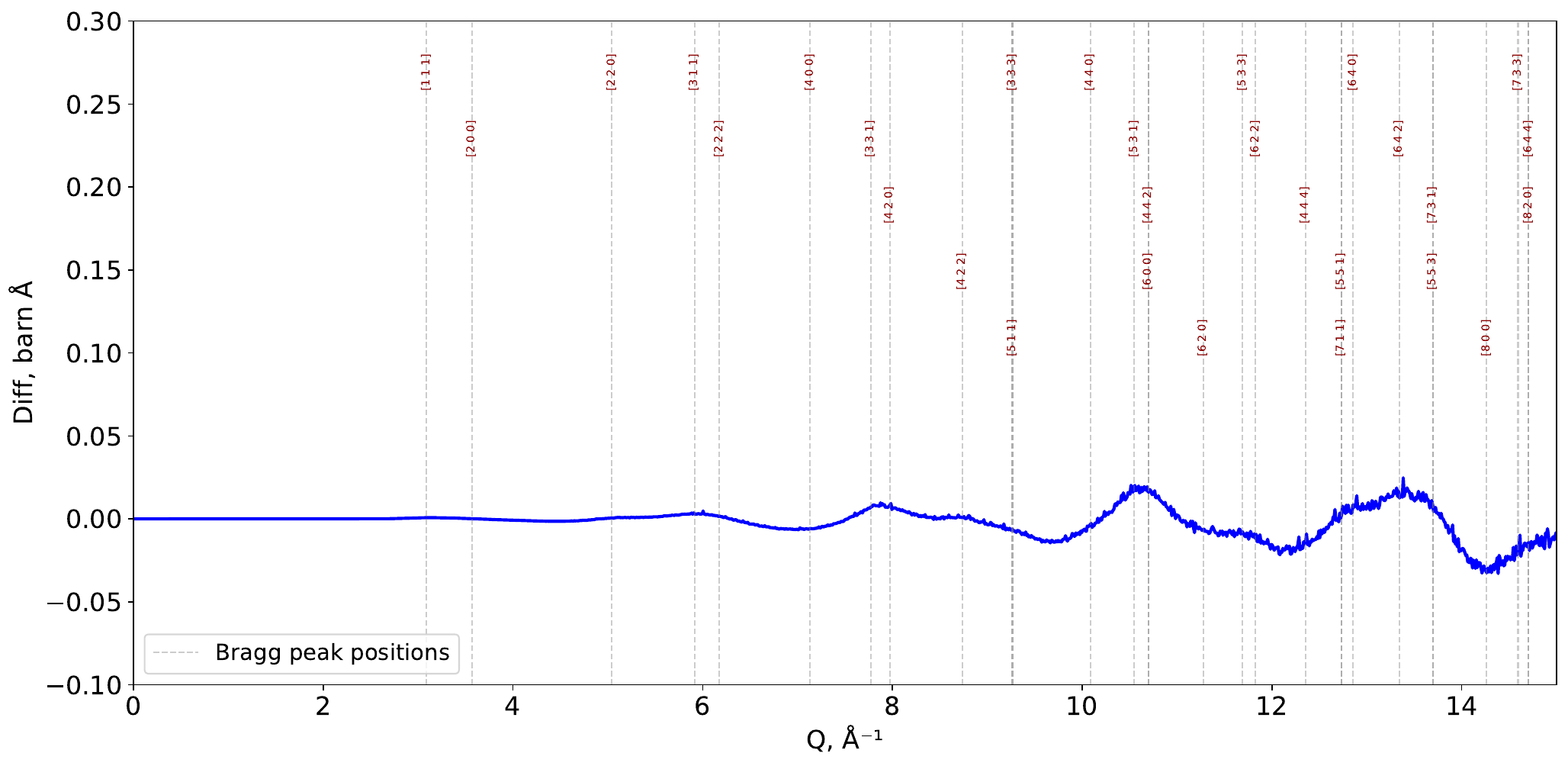}
    \caption{$M_0$ difference of two phonon scattering in Ni with Bragg peak positions}
    \label{fig:ni_300k_diff_analysis}
\end{figure}

\subsection{Differential scattering cross section}
\label{ssDiffXS}

The calculation of \texttt{BzScope} is also compared with an inelastic neutron scattering experiment~\cite{lih20k} performed on TOSCA~\cite{Colognesi2002} spectrometer at ISIS, as shown in Fig.~\ref{fig:lih_tosca}. The experimental sample is \text{LiH}, an alkali metal hydride, and measured at \SI{20}{\kelvin}, as described in section 3.4 of the article~\cite{BARRERA2005119}. The experiment result is actually the measurement of the differential cross section.

In order to obtain the calculated cross section, the contributions from both inelastic coherent single and two phonon events are obtained from \texttt{BzScope} calculation, and combine with the contributions from all other coherent and incoherent phonon events computed by NCrystal.

For BzScope, the crystal structure and force set of LiH (mp-23703) were downloaded and read from the online PhononDB. 
For OCLIMAX~\cite{Cheng2023}, the data for both the VDOS of phonon and the scattering intensity is captured directly from the pre-calculated OCLIMAX database~\cite{oclimaxdb}, with temperature set to \SI{0}{\kelvin}. The impact of the temperature difference between OCLIMAX and others can be ignored at this situation.

The calculated VDOS of phonon is shown in Fig.~\ref{fig:lih_vdos_sub}. Results from BzScope, OCLIMAX and Phonopy are compared. The VDOS from Phonopy is calculated with a mesh of $301 \times 301 \times 301$ at \SI{20}{\kelvin}. The VDOS from \texttt{BzScope} is calculated using the VEGAS algorithm at \SI{20}{\kelvin}. All of them are normalized by their areas. The result shows that there is good agreement between \texttt{BzScope} and Phonopy, except for energy at 12.9 THz and 24.1 THz, where BzScope's calculation has higher peak, while the result from OCLIMAX might not have converged.

The differential scattering cross section is shown in Fig.~\ref{fig:lih_xs_sub}. In order to compare with each other, the energy range is clamped to the common range. The results from both OCLIMAX and TOSCA are normalized by intensity calculated by BzScope. Although the computing results from both \texttt{BzScope} and OCLIMAX look like not well agree with that from experiment, the good consistency between \texttt{BzScope} and OCLIMAX is shown.
The inconsistency between calculation and experiment is mainly due to the open geometry of TOSCA, which brings complex background. Another possible impact may rise from the discrepancy of the crystal structure used in the calculation and in the experiment.

\begin{figure}[htbp]
  \centering
  \begin{subfigure}[b]{0.45\textwidth}
    \includegraphics[width=\textwidth]{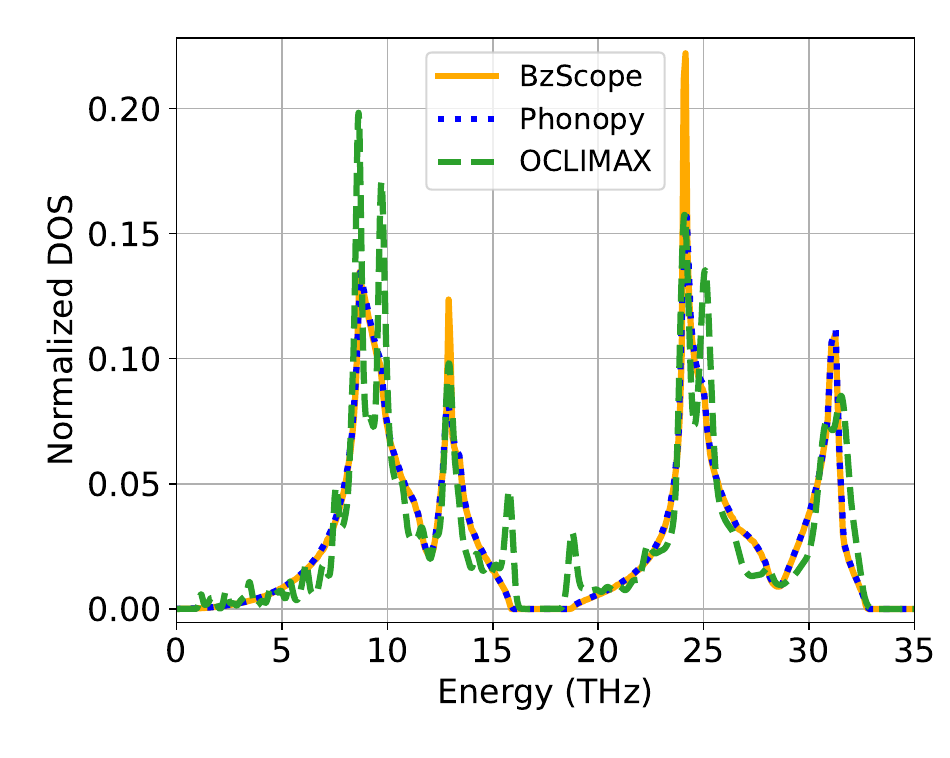}
    \caption{Computed VDOS}
    \label{fig:lih_vdos_sub}
  \end{subfigure}
  \hfill
  \begin{subfigure}[b]{0.45\textwidth}
    \includegraphics[width=\textwidth]{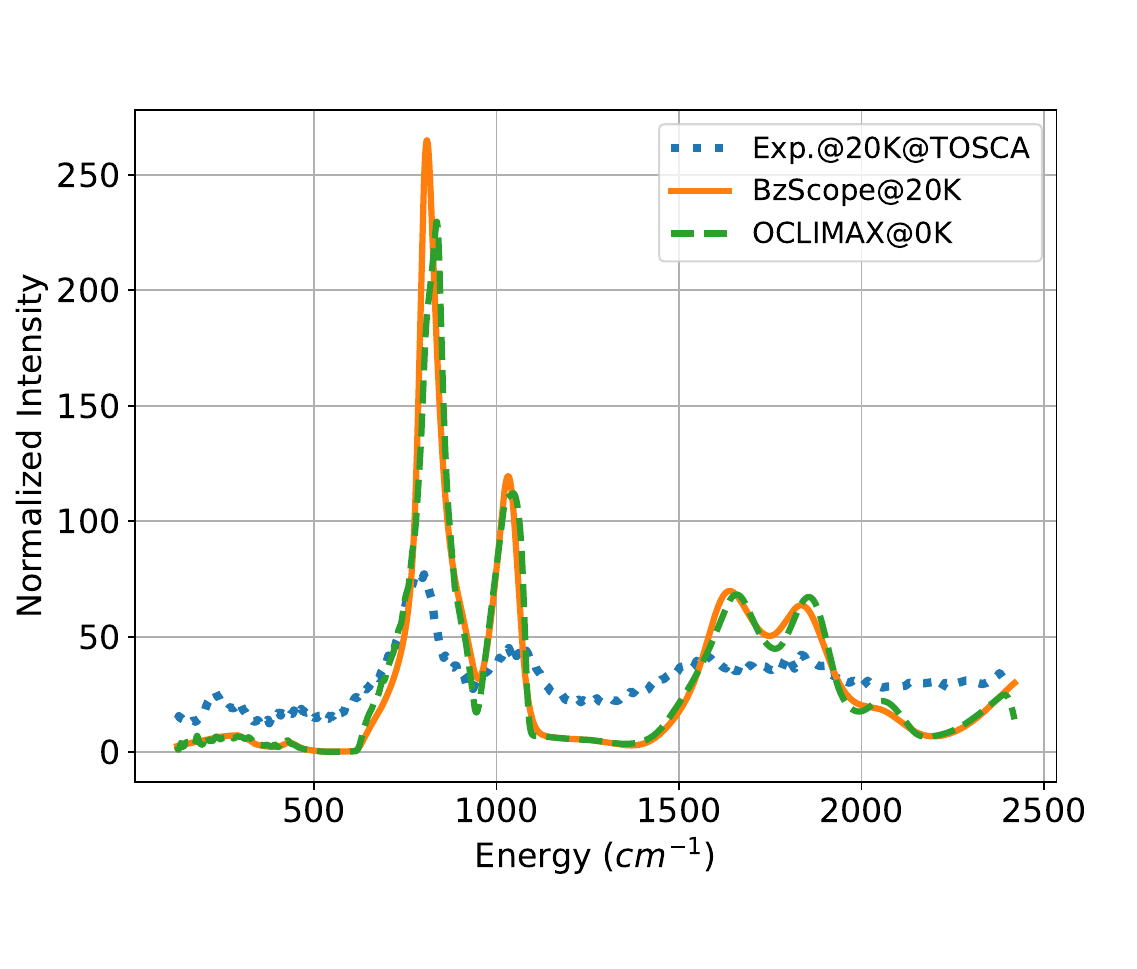}
    \caption{Differential cross section}
    \label{fig:lih_xs_sub}
  \end{subfigure}
  \caption{Differential scattering cross section of \( \text{LiH} \) at \SI{20}{\kelvin} from both calculation and experiment. Note the calculation of OCLIMAX takes place at \SI{0}{\kelvin}}
  \label{fig:lih_tosca}
\end{figure}

\subsection{Total cross section}
\label{ssTotXS}

NCrystal can take the external scattering functions into account via the \texttt{BzScope} plugin that is able to partially substitute the approximate inelastic scattering function in NCrystal.  
The \texttt{bz2nc} script is used in this section to export the material definition files that are compatible with the plugin. Hence the total cross section can be obtained from NCrystal to include additional elastic scattering and nuclear absorption processes. 
When experimental data from transmission measurements are relevant, the total cross section is calculated.
Otherwise, the integral cross section of inelastic scattering is calculated to concentrate to the scope of this work. 

The total cross sections of the two models are compared for Be at 300 K, along with the experimental data from EXFOR (11204003)~\cite{exfor11204003}, as shown in Fig.~\ref{fig:Be_total_300K}.
The \texttt{BzScope} exported data include coherent single and double phonon scattering functions.
Below the Bragg diffraction threshold, ca. \SI{5e-3}{\electronvolt}, the discrepancies between the models monotonically increase from zero to about 10\% at \SI{e-4}{\electronvolt}. Above that threshold, the Bragg diffraction is the dominant process, hence the discrepancies can not be observed. The decline of the first Bragg contribution in the experimental data is almost due to extinction effects, which are very prevalent in beryllium.
The \texttt{BzScope} curve agrees better with the experimental data, indicating that the \texttt{BzScope} model is more realistic.   

\begin{figure}[htbp]
  \centering
  \includegraphics[width=0.5\textwidth]{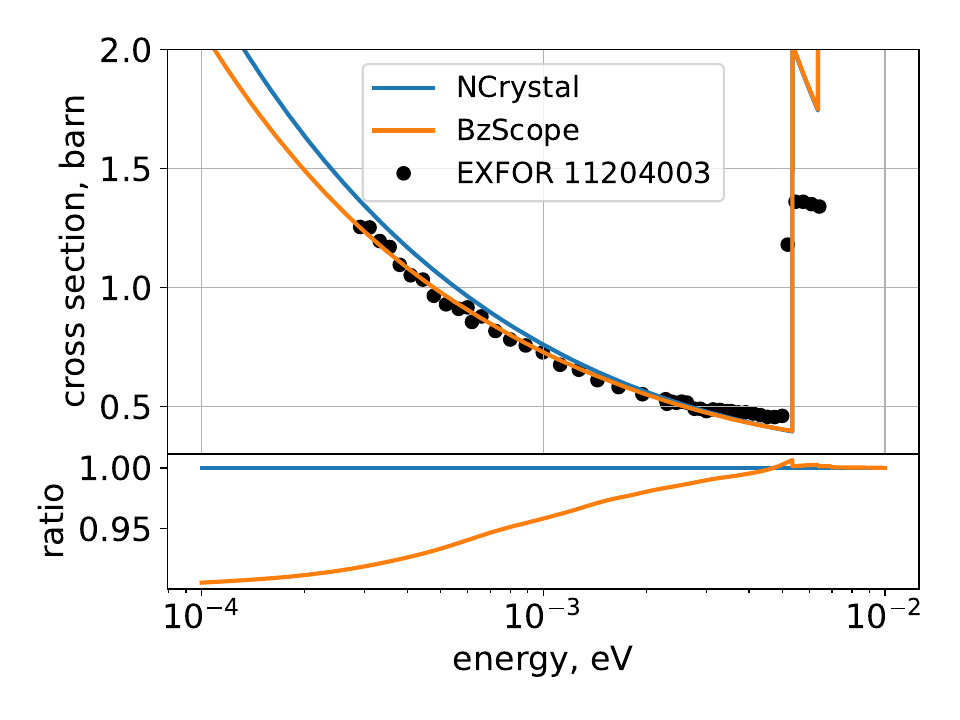}
  \caption{Total cross section of Be at \SI{300}{\kelvin}, coherent single and double phonon scattering functions are included in the \texttt{BzScope} curve. }
  \label{fig:Be_total_300K}
\end{figure}

\section{Conclusion}
Built upon the harmonic approximation and the adaptive Monte Carlo sampling policy, \texttt{BzScope} is capable of directly computing the contributions from incoherent single-phonon, coherent single-phonon and two-phonon processes to the scattering function for ideal powder crystalline materials. 
The calculations are numerically robust  across a wide range of neutron momentum transfers (up to $Q = 100~\text{\AA}^{-1}$) with high precision and resolution.

By adopting the Phonopy force constants format, \texttt{BzScope} can seamlessly integrate with various upstream DFT packages. Additionally, through the NCrystal \texttt{BzScope} plugin and the built-in bz2nc utility, \texttt{BzScope} can also be integrated with NCrystal to compute both differential and total cross sections on an absolute scale. Hence obtained scattering cross section can be used by downstream Monte Carlo softwares to simulate neutron scattering experiments.

Benchmark results show the calculations from \texttt{BzScope} are in good agreement with those from existing tools and experimental data, while also demonstrating its ability to reveal pair correlations between distinct atoms. These findings collectively confirm BzScope’s reliability.

Currently, \texttt{BzScope} supports only isotropic crystal powders. Support for single-crystal is under development. 
Magnetic neutron scattering is also being considered for the future versions.

\section*{Acknowledgements}
This work is supported by the National Key Research and Development Program of China (Grant No. 2022YFA1604100), the National Natural Science Foundation of China (Grant No. 12075266) and Guangdong Basic and Applied Basic Research Fund (Guangdong-Dongguan Joint Fund) (Grant No.2023A1515140057). 
The authors are grateful for valuable discussions and substantial support.


\bibliographystyle{elsarticle-num}
\bibliography{sample}


\end{document}